\documentclass[usenatbib]{mn2e}

\usepackage{graphicx}
\usepackage{amssymb}
\title[Chaos around black holes with discs or rings]
      {Free motion around black holes with discs or rings:\\
       between integrability and chaos --- III}
\author[P. Sukov\'a and O. Semer\'ak]
       {P. Sukov\'a\thanks{E-mail: lvicekps@seznam.cz}
        and
        O. Semer\'ak\thanks{E-mail: oldrich.semerak@mff.cuni.cz}\\
       Institute of Theoretical Physics,
       Faculty of Mathematics and Physics,
       Charles University in Prague,
       Czech Republic}
\begin{document}

\date{}

\pagerange{\pageref{firstpage}--\pageref{lastpage}} \pubyear{}

\maketitle

\label{firstpage}

\begin{abstract}
We continue the study of time-like geodesic dynamics in exact static, axially and reflection symmetric space-times describing the fields of a Schwarzschild black hole surrounded by thin discs or rings. In the first paper of this series, the rise (and decline) of geodesic chaos with ring/disc mass and position and with test particle energy was revealed on Poincar\'e sections and on time series of position or velocity and their power spectra. In the second paper we compared these results with those obtained by two recurrence methods, focusing on ``sticky" orbits whose different parts show different degrees of chaoticity. Here we complement the analysis by using several Lyapunov-type coefficients which quantify the rate of orbital divergence. After comparing the results with those obtained by the previous methods, we specifically consider a system involving a black hole surrounded by a small thin disc or a large ring, having in mind the configuration which probably occurs in galactic nuclei. Within the range of parameters which roughly corresponds to our Galactic center, we found that the black-hole accretion disc does not have a significant gravitational effect on the dynamics of free motion at larger radii, while the inner circumnuclear molecular ring (concentrated above 1 parsec radius) can only induce some irregularity in motion of stars (``particles") on smaller radii if its mass reaches 10 to 30\% of the central black hole (which is the upper estimate given in the literature), if it is sufficiently compact (which does not hold but maybe for its inner rim) and if the stars can get to its close vicinity. The outer dust ring between 60 and 100 parsecs appears to be less important for the geodesic dynamics in its interior.
\end{abstract}


\section{Introduction}

Black holes are the most conservative, today almost routine explanation of a whole bunch of high-energy astrophysical phenomena. Few doubt about their reality, but less certain is their accurate type. Models of accreting black holes standardly use Kerr metric for description of the gravitational field, disregarding the effect of the accreting (and any other) matter as well as possible non-stationarity and different global properties of the surrounding universe. While the main issue is still to confirm with certainty that black-hole horizons do occur in most galactic nuclei and in some X-ray binaries, it may soon become possible to test how well they correspond to what is described in general-relativity courses. It is thus desirable to discuss which effects can indicate deviations from the textbook ideal and what observational implications they may have.

One of almost unavoidable consequences of a departure from the Kerr geometry is a breakdown of the complete integrability of geodesic equations. This means, physically, that the dynamics of satellites freely orbiting astrophysical black holes should be ``weakly non-integrable" and possibly prone to chaos in some regions of the phase space \citep{Lukes-GAC-10}. In astrophysical systems with accreting black holes the matter elements have actually many reasons why to behave in a chaotic way, but we only focus on that resulting from {\em gravitational} influence of the accreting (or just surrounding) material. Since the latter is supposed to typically form a disc or a ring about the central black hole, it is natural to approximate the field of such a system by a suitable stationary and axially symmetric (and orthogonally transitive) space-time. Though at least outside of the sources (in a vacuum) the Einstein equations are known to be completely integrable in such a case (they are usually written and treated in the form of the Ernst equation) and ``all" such space-times have actually been covered within wide families of solutions, mostly obtained by ``solution-generating" techniques, these have not proved successful in describing physically reasonable fields, if having an interpretation at all.

The only case when the Einstein equations allow for a description of the composite-source field (for ``superposition") which is practical for actual computations is the {\em static} and axially symmetric situation. Then the metric contains just two unknown functions, of which the first satisfies Laplace equation (thus adds up linearly) and the second one is obtained from the first by a suitable line integral (see e.g. \citealt{SemerakS-10} for more detailed introduction and references). Needless to say, a static setting excludes rotation, which is a serious limitation concerning that the black-hole accretion systems are expected to bear a considerable angular momentum. On the other hand, the gravitational implications of rotation --- the effects of rotational dragging --- fall very quickly with distance from the source, so at least for satellites orbiting not so close to the centre one can assume they do not play an important role.\footnote
{However, in the long-term dynamics, it is difficult to estimate a priori which of the tiny effects will bring more important ``perturbation", so one should definitely try to incorporate rotation in some way, if only to learn whether chaos tends to be enhanced or suppressed by the dragging effects.}

It is a question whether some of the astrophysical black-hole--dominated systems are ``clean" enough, so that the restriction to a pure gravity-driven orbital motion can be adequate. This is hardly so in the case of X-ray binaries, whose radiation itself indicates that the components interact strongly and that a corresponding amount of matter and/or fields has to be present; moreover, the geometry may be rather far from static (and even stationary) and axially symmetric in these systems. The same can generically be estimated in highly active galactic nuclei. On the other hand, in less active nuclei, with lower density of interstellar matter, the orbiting of stars around the central supermassive black hole could be well approximated by geodesic motion in the field of the hole perturbed by an accretion disc or/and by a circum-nuclear structures of colder material on larger radii.

Let us recall that in the first paper \citep{SemerakS-10} of this series (see also \citealt{Sukova-11}) we considered the system of an (originally) spherically symmetric black hole surrounded by a thin disc or a ring and tried to learn how the dynamics of time-like geodesics in its field depends on parameters. More specifically, we placed the inverted 1st (and also 4th) member of the Morgan-Morgan counter-rotating disc family or the Bach-Weyl ring around the Schwarzschild hole in a concentric manner and were observing the rise (and decline) of geodesic chaos with disc/ring mass and position and with test-particle energy, as revealed on Poincar\'e sections and on time series of position or velocity and their power spectra. In the second paper \citep{SemerakS-12} (see also \citealt{SukovaS-12}) we compared the results with those obtained by two simple and powerful recurrence methods, the method based on averaging of directions in which the system passes through a pre-defined phase-space cells, and the method of recurrence plots based on statistics over the recurrences to these cells themselves. We mainly focused there on ``sticky" orbits whose different parts show different degrees of chaoticity, because such orbits, lying just between regular and strongly chaotic regime, offer the best chance to test different methods and their sensitivity to different dynamical features.

We refer to the preceding two papers for general introduction including a number of references and for details (which will not be repeated here). Let us just remind that the system we consider is fully characterised by the black-hole mass $M$, by the disc/ring mass ${\cal M}$ and by the disc inner radius $r_{\rm disc}$ or the ring radius $r_{\rm ring}$ (the radial coordinate $r$ is of Schwarzschild type). The main parameters (constants) of geodesic motion are specific energy at infinity ${\cal E}\equiv -u_t$ and azimuthal angular momentum at infinity $\ell\equiv u_\phi$ ($u_\mu$ is the covariant four-velocity). The Poincar\'e sections show transits of the geodesics through the equatorial plane (the plane where the disc or the ring resides) in the $(r,u^r)$ axes; regarding that the system is reflectionally symmetric with respect to the equatorial plane, we record passages in both directions. We use geometrised units in which $c=G=1$.

The present paper extends the analysis by computing several Lyapunov-type coefficients which quantify the rate of orbital divergence, namely the maximal Lyapunov characteristic exponent and the computationally more suitable FLI and MEGNO indicators (they are introduced in section \ref{coefficients}). After comparing the results with those obtained by the other methods previously (section \ref{comparison-with-previous}), we specifically consider a system involving a black hole surrounded by a small thin ``accretion" disc or a large ring (approximating the ``cold" torus), having in mind the configuration which probably occurs in galactic nuclei (section \ref{galactic-nuclei}). If choosing the parameters according to the values observed in our Galaxy, it turns out that the innermost accretion disc around the central black hole does not significantly affect the dynamics of free motion on larger radii, whereas the inner circumnuclear molecular ring can make some of the lower-radius orbits chaotic, provided that its mass is sufficiently large and compact enough and that the stars can approach it closely enough.

\subsection{Reminding the metrics considered}

The formulas describing the exact superposition of a vacuum static and axially symmetric (originally Schwarzschild) black hole with a concentric thin disc or ring were given in previous papers of this series (together with the original literature), so we will only list them very shortly for reference.
In a vacuum with the above space-time symmetries, the metric can be put into the Weyl form
\begin{equation}  \label{Weylmetric}
  {\rm d}s^2=-e^{2\nu}{\rm d}t^2
             +\rho^2 e^{-2\nu}{\rm d}\phi^2
             +e^{2\lambda-2\nu}({\rm d}\rho^2+{\rm d}z^2),
\end{equation}
where $(t,\rho,z,\phi)$ are Weyl coordinates (of cylindrical type), $t$ and $\phi$ representing Killing time and azimuthal coordinate and $\rho$ and $z$ covering the meridional planes. The two unknown functions $\nu$, $\lambda$ only depend on $\rho$ and $z$; the ``gravitational potential" $\nu$ satisfies the Laplace equation, so it superposes linearly, while $\lambda$ is found by quadrature
\begin{equation}  \label{lambda}
  \lambda=
  \int_{\rm axis}^{\rho,z}
      \rho\left\{\left[(\nu_{,\rho})^2-(\nu_{,z})^2\right]{\rm d}\rho
                 +2\nu_{,\rho}\nu_{,z}{\rm d}z\right\},
\end{equation}
computed along any line within the vacuum region (note that $\lambda=0$ on vacuum parts of the axis).

The Schwarzschild field (of mass $M$) is described by
\begin{eqnarray}
  \nu_{\rm Schw}&=&\frac{1}{2}\,\ln\frac{d_1+d_2-2M}{d_1+d_2+2M}
                 = \frac{1}{2}\,\ln\left(1-\frac{2M}{r}\right),\\
  \lambda_{\rm Schw}&=&\frac{1}{2}\,\ln\frac{(d_1+d_2)^2-4M^2}{4\Sigma}
                     = \frac{1}{2}\,\ln\frac{r(r-2M)}{\Sigma}\;,
  \label{lambdaSchw}
\end{eqnarray}
where
\begin{eqnarray*}
  d_{1,2}&\equiv&\sqrt{\rho^2+(z\mp M)^2}=r-M\mp M\cos\theta \,,\\
  \Sigma\equiv d_1 d_2
        &=&\sqrt{(\rho^2+z^2+M^2)^2-4z^2 M^2} \\
        &=&(r-M)^2-M^2\cos^2\theta \,.
\end{eqnarray*}
The second expressions are in Schwarzschild coordinates $(r,\theta)$ which are related to the Weyl coordinates by
\begin{eqnarray*}
\lefteqn{
  \rho=\sqrt{r(r-2M)}\,\sin\theta\,, \quad
  z=(r-M)\,\cos\theta \,;} \\
\lefteqn{
  r-M=\frac{1}{2}(d_2+d_1)\,, \quad
  M\cos\theta=\frac{1}{2}(d_2-d_1) \,.}
\end{eqnarray*}
The Schwarzschild-type coordinates (or other spheroidal coordinates like the isotropic ones) are actually more natural for space-times containing a black hole, because the latter's horizon is represented, at given Killing time $t$, as a sphere ($r=2M$) in them, whereas in the Weyl coordinates it is a finite part of the symmetry axis ($\rho=0$, $|z|\leq M$). Outside of the thin source with no radial pressure, the complete metric reads then
\begin{eqnarray}
  {\rm d}s^2=
    &-&\!\!\!\!\!\left(1-\frac{2M}{r}\right)e^{2\hat\nu}{\rm d}t^2
      +\frac{e^{2\hat\lambda-2\hat\nu}}{1-\frac{2M}{r}}\;{\rm d}r^2 \nonumber \\
    &+&\!\!\!\!r^2 e^{-2\hat\nu}
       (e^{2\hat\lambda}{\rm d}\theta^2+\sin^2\theta\,{\rm d}\phi^2),
\end{eqnarray}
where
$\hat\nu(r,\theta)$ is the potential of the external source and
$\hat\lambda(r,\theta)\equiv\lambda-\lambda_{\rm Schw}$ with $\lambda_{\rm Schw}$ given by (\ref{lambdaSchw}).

We consider superpositions with the inverted 1st member of the Morgan-Morgan counter-rotating thin-disc family and with the Bach-Weyl thin ring.
The disc potential reads
\begin{equation}
  \nu_{\rm iMM1}
  = -\frac{{\cal M}}{\pi(\rho^2+z^2)^{3/2}}
     \left(P_1\,{\rm arccot}\,S-P_2 S\right)
\end{equation}
in Weyl coordinates, where
\begin{eqnarray*}
  P_1&\equiv& 2\rho^2+2z^2-b^2\,\frac{\rho^2-2z^2}{\rho^2+z^2} \;, \\
  P_2&\equiv& \frac{1}{2}\left(3\Sigma-3b^2+\rho^2+z^2\right) \,, \\
  S  &\equiv& \sqrt{\frac{\Sigma-\rho^2+b^2-z^2}{2\,(\rho^2+z^2)}} \;,
\end{eqnarray*}
${\cal M}$ and $b$ being mass and Weyl inner radius of the disc,
and $\Sigma\equiv\sqrt{(\rho^2-b^2+z^2)^2+4b^2 z^2}$ now.
The ring potential reads
\begin{equation}
  \nu_{\rm BW}=-\frac{2{\cal M}K(k)}{\pi l_2}\;,\\
\end{equation}
where ${\cal M}$ and $b$ are again mass and Weyl radius of the ring,
$K(k)\equiv\int_0^{\pi/2}\frac{{\rm d}\phi}{\sqrt{1-k^2\sin^2\phi}}$
is the complete Legendre elliptic integral of the 1st kind,
$k'^2=\frac{(l_1)^2}{(l_2)^2}\,$,
$k^2=1-k'^2=\frac{4\rho b}{(l_2)^2}\,$, and
$l_{1,2}\equiv\sqrt{(\rho\mp b)^2+z^2}\,$.

\section{Quantifying the orbital divergence}
\label{coefficients}

One of the basic symptoms of chaos is a quick divergence of phase trajectories in certain directions --- the well-known sensitive dependence on initial conditions. The rate of this tendency can be quantified by several coefficients whose main purpose is to distinguish between polynomial and exponential divergence. Their computation typically involves a sequence of evolutions and renormalisations of a certain relative position vector, which requires a choice of {\em the} time coordinate. This is of course not covariant in principle, and may even be practically ambiguous if space-time is complicated enough. Fortunately, when the space-time has a time-like symmetry (it is stationary), there does exist a privileged global time which usually is a reasonable choice for computations. Also, in agreement with the knowledge and experience that the nature of dynamics is not sensitive to the metric used in the phase space, it was shown that this also applies to the most important Lyapunov-type exponents; as quoted from \cite{GelfertM-10}: ``our results show once and for all that Lyapunov exponents, entropies, and dimension-like characteristics can be used to make invariant assertions about chaos. However, the same results also show that the {\em values} of some quantities that have been previously conjectured to be invariant, such as the information dimension and topological entropy, are not invariant in general."

In this paper, we focus on three quantities which measure the rate of orbital divergence: the Lyapunov characteristic exponents (LCEs), the fast Lyapunov indicator (FLI) and the mean exponential growth of nearby orbits (MEGNO).
(See e.g. \citealt{MaffioneDCG-11} for a comparison of these quantities with several other indicators derived from deviation vectors.)
The LCEs were introduced by Lyapunov in 1892 (see, e.g., \citealt{BenettinGGS-80}) and are mostly computed in two different ways, namely by solving the variational equations along with the equations of motion of the given dynamical system, or by following the evolution of separation of its two nearby orbits. The variational approach is generally more accurate and reliable, but for relativistic systems it usually involves complicated equations that are very difficult to integrate; the two-particle method can be an efficient and much more convenient option which however has to be employed with caution \citep{TancrediSR-01,Wu-06}. We here follow the procedure proposed by \cite{Wu-03,Wu-06}.

The LCEs describe the rate of orbital divergence in the neighbourhood of a given trajectory. Considering the initial conditions distributed on a sphere of small radius ${\bf \Delta w}$ around some given point ${\bf w}$, the orbital flow deforms the sphere into an ellipsoid whose half-axes evolve according to ${\rm e}^{\lambda_i t}\Delta w_i$, where $\lambda_i$ is the LCE in the direction of the $i$-th axis and $\Delta w_i$ is the respective component of ${\bf \Delta w}$. The orbits thus diverge exponentially in certain direction if the corresponding $\lambda_i$ is positive. Though the information about phase dynamics is encoded in the whole spectrum of LCEs, it is thus most important to determine the {\em maximal} LCE (mLCE) (we will call it $\lambda_{\rm max}$) in order to decide whether the system is chaotic or not. Actually, for a randomly selected two nearby trajectories the separation vector always has some nonzero component in the direction corresponding to the mLCE and after a short time this component outweighs the others; denoting such two orbits by ${\bf w}(t)$ and ${\bf w}'$(t), the mLCE is given by the limit
\begin{equation}  \label{mlce_cl}
  \lambda_{\rm max}=
  \lim_{t\to\infty} \frac{1}{t}\,\ln \frac{\Vert{\bf \Delta w}(t)\Vert}{\Vert{\bf \Delta w}(0)\Vert} \;,
\end{equation}
where $\Vert{\bf \Delta w}(t)\Vert\equiv\Vert{\bf w}(t)-{\bf w'}(t)\Vert$ is the norm of the displacement vector in the phase space.

LCEs belong to the quantities which do bring invariant information (it is their sign in this case) but whose {\em value} is coordinate dependent \citep{GelfertM-10}. In relativity, the main issue is the choice of time $t$. In stationary space-times, there exists a privileged, Killing time which offers a natural option for a study of test-particle dynamics. However, \cite{Wu-06} argued that one should examine, irrespectively of the type of space-time, the evolution of {\em proper} distance between the orbits in {\em proper} time rather than coordinate quantities. They also claim (i) that it is sufficient to follow this evolution in the configuration space (i.e. regardless of the momentum dimensions) in order to distinguish between regular and chaotic regions, and (ii) that the displacement vector need not be projected on a certain space-like hypersurface. Hence the formula
\begin{eqnarray}
  \tilde{\lambda}_{\rm max} &=&
  \lim_{\tau\to\infty} \frac{1}{\tau}\,
  \ln\frac{|\Delta{\bf x}(\tau)|}{| \Delta {\bf x}(0)|} \,, \label{mlce} \\
  |\Delta{\bf x}(\tau)| &\equiv& \sqrt{|g_{\mu\nu}\Delta x^\mu\Delta x^\nu|(\tau)} \;,
\end{eqnarray}
where proper time $\tau$ is the independent integration variable and $\Delta{\bf x}$ is the separation vector whose norm represents the momentary proper distance between the two neighbouring orbits in the configuration space. In the infinite-time limit this more convenient way of computation should coincide with the classical definition of mLCE. We have not checked this statement in detail, but will follow it, not even emphasising it in notation, so we will suppose $\tilde{\lambda}_{\rm max}\equiv\lambda_{\rm max}$ and omit the tilde hereafter. In order to keep in focus the orbital flow in the vicinity of the reference trajectory, the separation vector has to be renormalised whenever it reaches a certain  prescribed value; together with the latter, the velocity deviation vector ${\bf \Delta u}(\tau)={\bf u}(\tau)-{\bf u'}(\tau)$, where ${\bf u}$ is the four-velocity tangent to the reference world-line, has to be renormalised by the same factor $|{\bf \Delta x}(0)|/|{\bf \Delta x}(\tau)|$.

The main disadvantage of the original LCEs is their very slow convergence to the final value, which means that for weakly chaotic systems a very long integration time is often necessary to prove the nature of their orbits. Therefore, a number of related quantifiers of orbital deviation has been proposed whose computation converges faster and which thus reveal the nature of orbits in a significantly shorter integration time.
\cite{FroeschleLG-97} suggested the so called fast Lyapunov indicator (FLI) which is tied very straightforwardly to the idea of exponential versus polynomial grow of the separation vector: it is related to its norm by
\begin{equation}
  {\rm FLI}(t) = \log_{10} \frac{\Vert{\bf \Delta w}(t)\Vert}{\Vert{\bf \Delta w}(0)\Vert} \;,
\end{equation}
or, when restricting only to its configuration-space part as above,
\begin{equation}  \label{FLI_def}
  {\rm FLI}(\tau) = \log_{10} \frac{|{\bf \Delta x}(\tau)|}{|{\bf \Delta x}(0)|} \;.
\end{equation}
FLI($\tau$) grows considerably faster for chaotic than for regular trajectories and this trend is evident much earlier than $\lambda_{\rm max}$ approaches its limit value. The degree of chaoticity of different phase-space layers can thus be effectively compared according to the values of FLI($\tau_{\rm max}$) found for their orbits (for some suitably chosen $\tau_{\rm max}$), even though these values do not have an invariant meaning.

Classification of dynamics according to the rate of growth of some quantifier (FLI) is practical for a small number of orbits, but for an extensive automatic survey which should scan a large part of the phase space it is more desirable to process just {\em value} of some quantity, if possible a time-independent one. Such kind of quantity was proposed by \cite{Cincotta-00}: the mean exponential growth factor of nearby orbits (MEGNO), computed as the mean temporal moment of the time change of $\ln\delta(t)$, where $\delta(t)$ is the variation of trajectory, namely
\begin{equation}
  Y(t) = \frac{2}{T}\int_0^T \frac{\dot{\delta}(t)}{\delta(t)}\,t\,{\rm d}t \;.
\end{equation}
The authors showed that for regular orbits the value of MEGNO tends to~2 with an additional bounded oscillating term, whereas it grows linearly for chaotic orbits with a slope corresponding to the value of mLCE ($Y(\tau)\approx\lambda_{\rm max}\tau$ for large enough proper time); for a sufficiently long interval $T$ the values clearly distinguish between the two regimes. Not long ago \cite{MestreCG-11} found an analytic relation between FLI and MEGNO,\footnote
{When comparing this paper with that by \cite{Cincotta-00}, mind the different definitions of FLI, with decadic logarithm in the older paper whereas with natural logarithm in the more recent one. This brings the $\ln(10)$ factor into the following relation.}
\begin{equation}  \label{MEGNO_def}
  Y(\tau) = 2\,[{\rm FLI}(\tau)-\overline{\rm FLI}(\tau)]\,\ln(10) \,,
\end{equation}
where the FLI time average is given by
\begin{equation}
  \overline{\rm FLI}(\tau)=\frac{1}{\tau}\int_0^\tau {\rm FLI}(s)\,{\rm d}s \,.
\end{equation}
Since the value of MEGNO oscillates with the same frequency as that of FLI, we can compute its time average in the same way,
\begin{equation}  \label{prum_MEGNO}
  \overline{Y}(\tau)=\frac{1}{\tau}\int_0^\tau Y(s)\,{\rm d}s \,.
\end{equation}
The average MEGNO $\overline{\rm Y}(\tau)$ behaves smoothly and for $\tau$ long enough clearly distinguishes between regular and chaotic motion, hence it is a most suitable indicator for automatic computations and can be easily determined from the course of FLI. Furthermore, computing the linear regression for MEGNO($\tau$) one infers the value of mLCE. In the literature, the MEGNO indicator has been mainly employed to study galactic and planetary dynamics. The FLI has also been computed for general-relativity problems (motion in black-hole fields), see \cite{Han-08a,Han-08b}.

In the next section, we compute the mLCE, FLI and MEGNO coefficients for our system of geodesics in the field of a static and axially symmetric (originally Schwarzschild) black hole surrounded concentrically by an inverted 1st Morgan-Morgan counter-rotating thin disc. The sources are uniquely specified by the relative disc mass ${\cal M}/M$ and by the disc's inner Schwarzschild radius $r_{\rm disc}$. The geodesics are characterised by their conserved energy and angular momentum at infinity per unit rest mass, ${\cal E}$ and $\ell$; the role of another ``integral of motion" is played by four-velocity normalisation $g_{\mu\nu}u^\mu u^\nu=-1$. The remaining freedom is exploited, in order to scan the phase space properly, by launching the particles from different points within a certain accessible domain and in different local directions.
We will compare the information captured by the coefficients with each other as well as with results obtained by different methods in previous papers \citep{SemerakS-10,SemerakS-12}.

\begin{figure*}
\includegraphics[width=\textwidth]{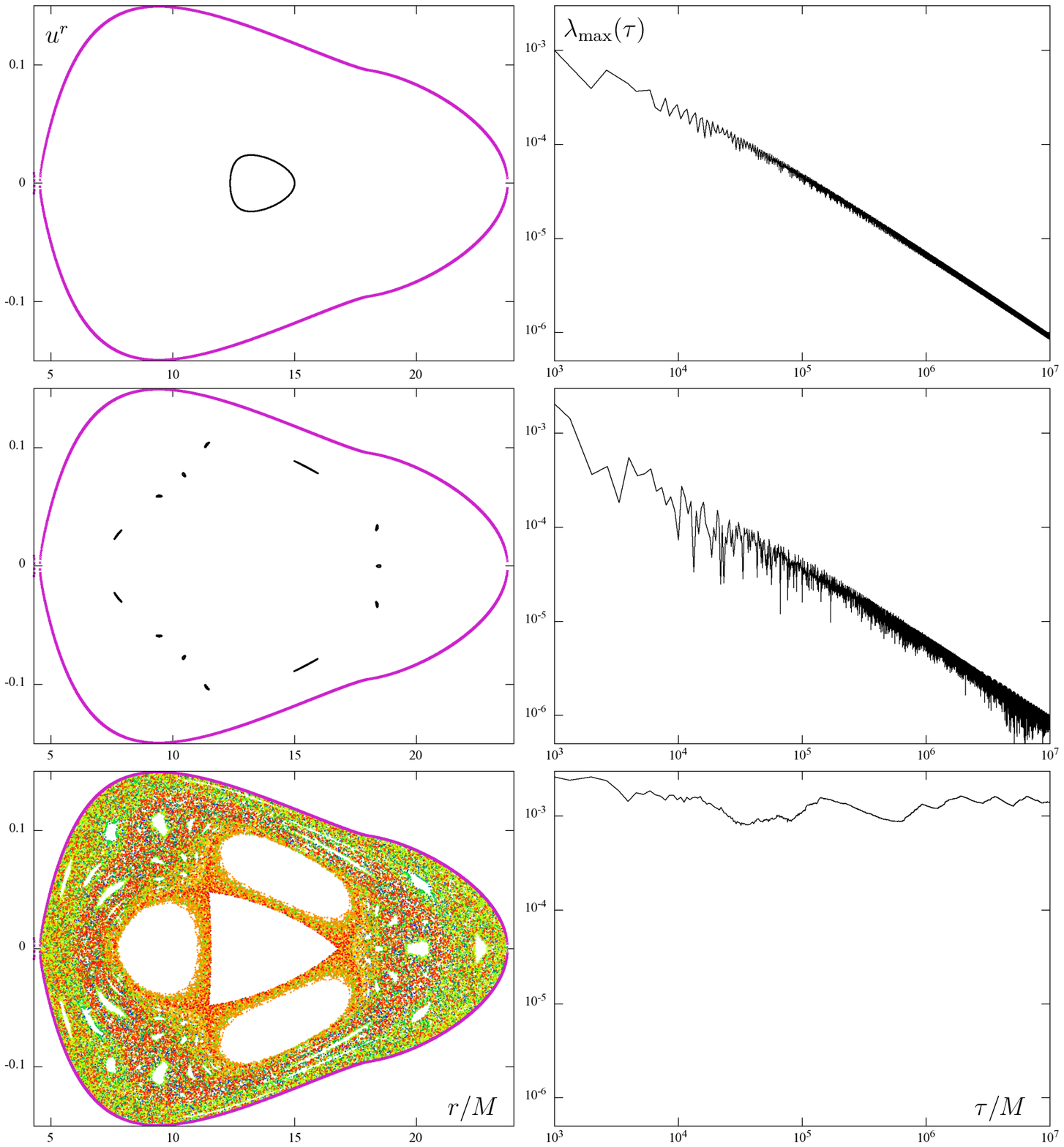}
\caption
{Poincar\'e sections $\theta=\pi/2$ (left) and maximal Lyapunov characteristic exponents $\lambda_{\rm max}(\tau)$ (mLCEs, right) for two regular (A, B) and one chaotic (C) geodesic in the field of a black hole surrounded by the inverted 1st Morgan-Morgan disc. The disc parameters are ${\cal M}=M/2$, $r_{\rm disc}=18M$, and the specific energy and angular momentum of the geodesics are ${\cal E}\equiv -u_t=0.9532$, $\ell\equiv u_\phi=3.75M$. The last Poincar\'e section is coloured according to the proper time of the passages; the time increases from blue to red in the order of the visible-spectrum colours. Different colour tones indicate that the orbit is not distributed uniformly over the chaotic layer, but rather lingers in a specific volume for some time and then ``switches" to a different one. The obtained values of $\lambda_{\rm max}(\tau_{\rm max})$ are $(9.35\pm 0.38)\cdot 10^{-7}M^{-1}$ for the upper row (orbit A), $(8.78\pm 0.58)\cdot 10^{-7}M^{-1}$ for the middle row (orbit B) and $(1.405\pm 0.007)\cdot 10^{-3}M^{-1}$ for the bottom row (orbit C). The FLI and MEGNO values for the same orbits are plotted in figure \ref{fig-4}.
\label{fig-1}
}
\end{figure*}

\begin{figure*}
\includegraphics[width=\textwidth]{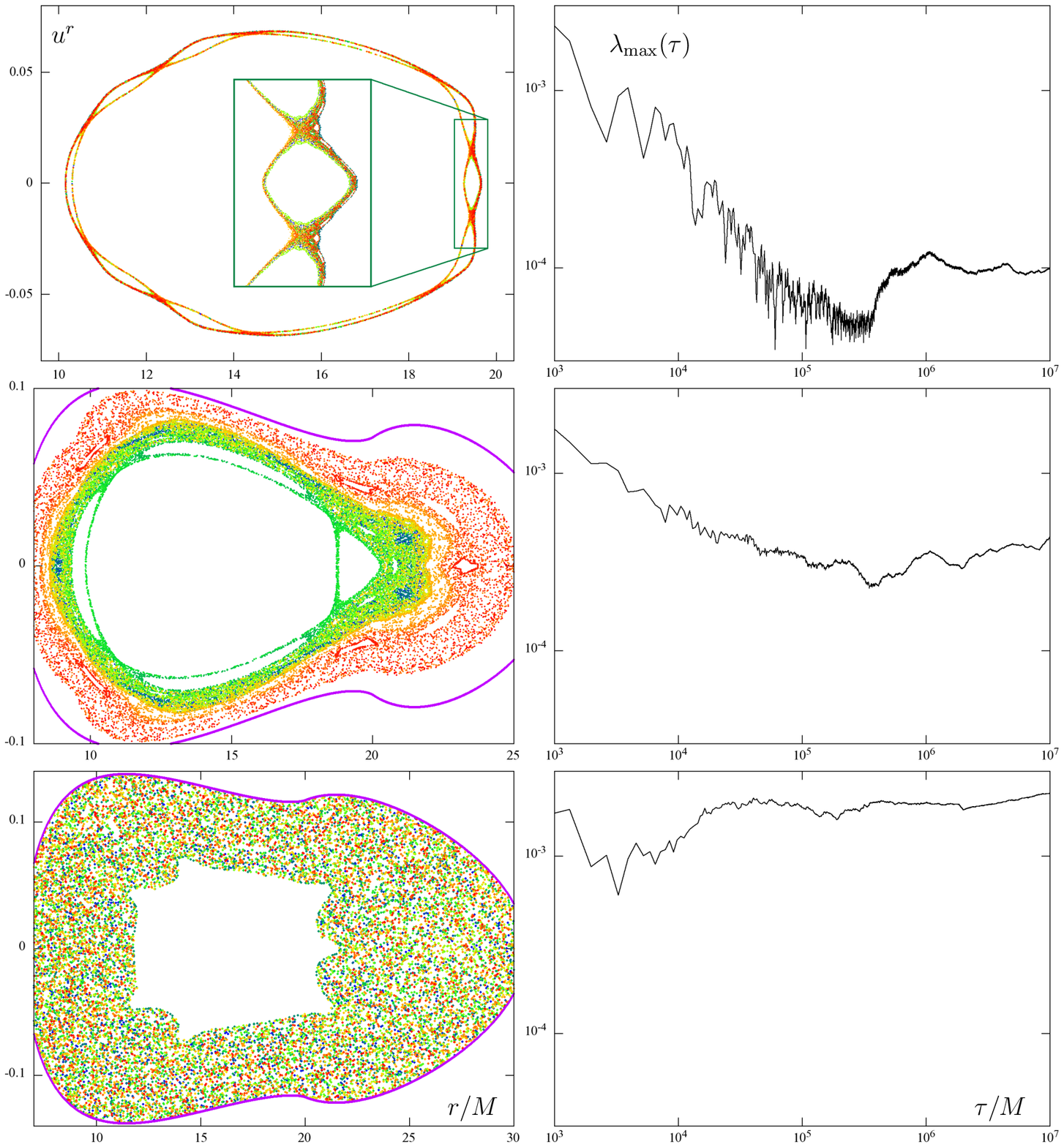}
\caption
{Poincar\'e sections (left) and mLCEs $\lambda_{\rm max}(\tau)$ (right) for three chaotic orbits (D, E, F) of different degrees of irregularity. The global parameters are ${\cal M}=0.94M$, $r_{\rm disc}=20M$, ${\cal E}=0.947$, $\ell=4M$ in the upper row (D); ${\cal M}=1.3M$, $r_{\rm disc}=20M$, ${\cal E}=0.9365$, $\ell=4M$ in the middle row (E); and ${\cal M}=1.3M$, $r_{\rm disc}=20M$, ${\cal E}=0.941$, $\ell=4M$ in the bottom row (F). The orbits clearly fill phase-space layers of different volumes, in agreement with the obtained values $\lambda_{\rm max}(\tau_{\rm max})=(9.88\pm 0.04)\cdot 10^{-5} M^{-1}$ (up, D), $\lambda_{\rm max}(\tau_{\rm max})=(4.28\pm 0.06)\cdot 10^{-4}M^{-1}$ (middle, E) and $\lambda_{\rm max}(\tau_{\rm max})=(2.250\pm 0.004)\cdot 10^{-3}M^{-1}$ (bottom, F). Colouring of the passages through the Poincar\'e surface by proper time (it increases in the order blue $\rightarrow$ green $\rightarrow$ yellow $\rightarrow$ red) reflects that the 2nd trajectory lingers for a longer time near the periodic island and only then sets off into the outer part of the phase space, whereas the 3rd trajectory is rather uniformly distributed across the layer for all the time. Yet in a certain interval of time (around $\tau\sim 2000M$) the mLCE of the third orbit is {\em smaller} than the mLCE of the second one. The orbits plotted here also appear in the animations which are accessible in the online version. The same orbits D, E, F are also processed in figures \ref{fig-3} and \ref{fig-5}.
\label{fig-2}
}
\end{figure*}

\begin{figure*}
\includegraphics[width=\textwidth]{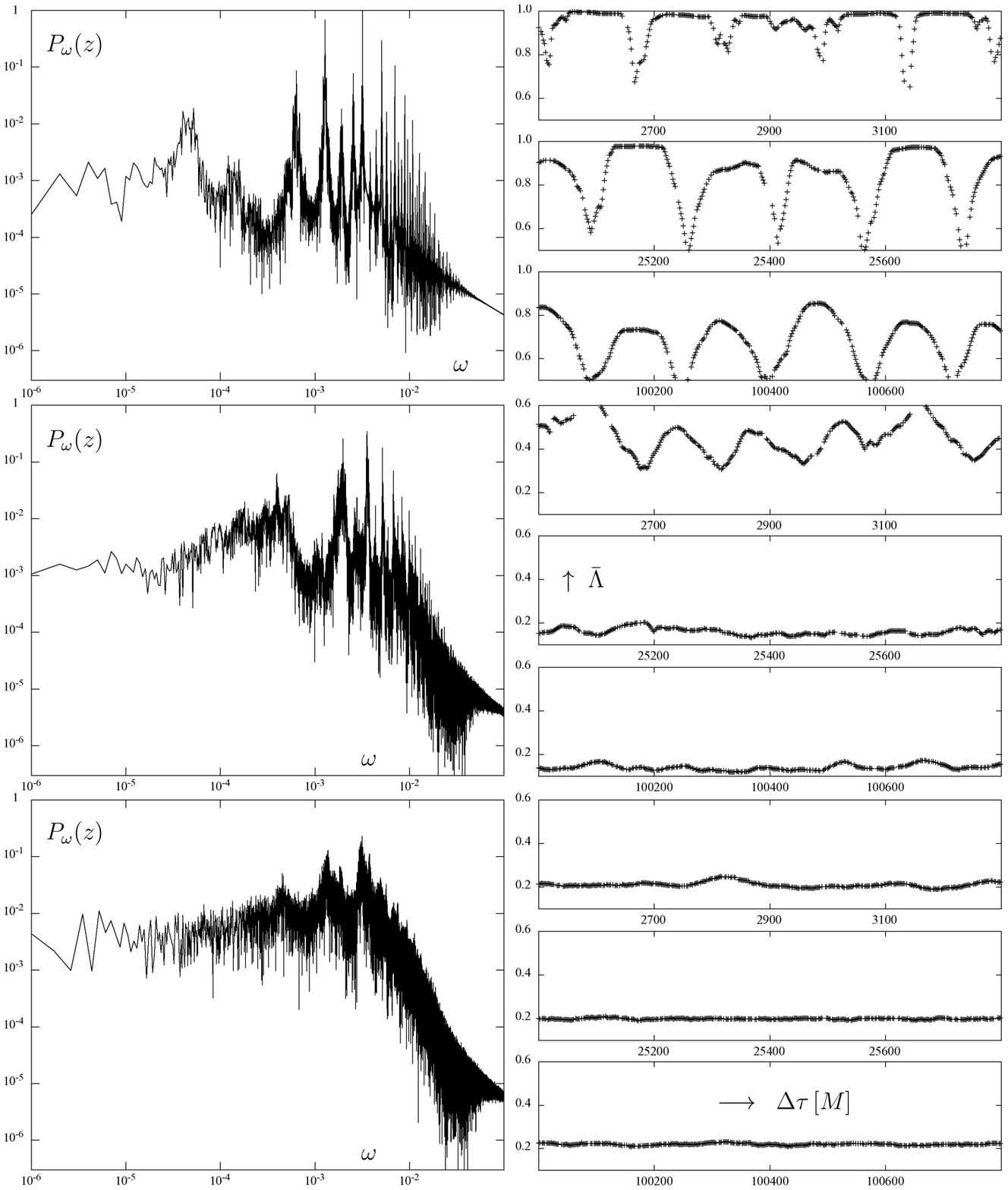}
\caption
{Power spectra of vertical position $z=r\cos\theta$ (left column) and average of the directions in which a given geodesic recurs to a prescribed phase-space mesh, $\bar\Lambda$ (right column), plotted for the same three orbits (D, E, F) whose Poincar\'e diagrams and mLCEs have been shown in figure \ref{fig-2}. In contrast to figure \ref{fig-2} where the orbits were computed up to $\tau_{\rm max}=10^7 M$, here only their parts reaching to $\tau_{\rm max}=10^6 M$ have been processed. The Kaplan-Glass indicator $\bar\Lambda$ decreases in general (from 1 and the more the less regular is the orbit) with the time shift $\Delta\tau$ used to reconstruct a phase space from a given time series; the $\bar\Lambda(\Delta\tau)$ dependence for all three orbits is drawn for three separate time-shift intervals, (2500--3300)$M$, (25000--25800)$M$ and (100\,000--100\,800)$M$. The first orbit is just very weakly chaotic in comparison with the other two. See section \ref{comparison-with-previous} for more discussion. The FLI and MEGNO values for the same orbits are plotted in figure \ref{fig-5}. Note that vertical-axes ranges ($\bar\Lambda$) are different on the right, 0.5--1.0 for the top orbit whereas 0.1--0.6 for the remaining two.
\label{fig-3}
}
\end{figure*}

\begin{figure*}
\includegraphics[width=\textwidth]{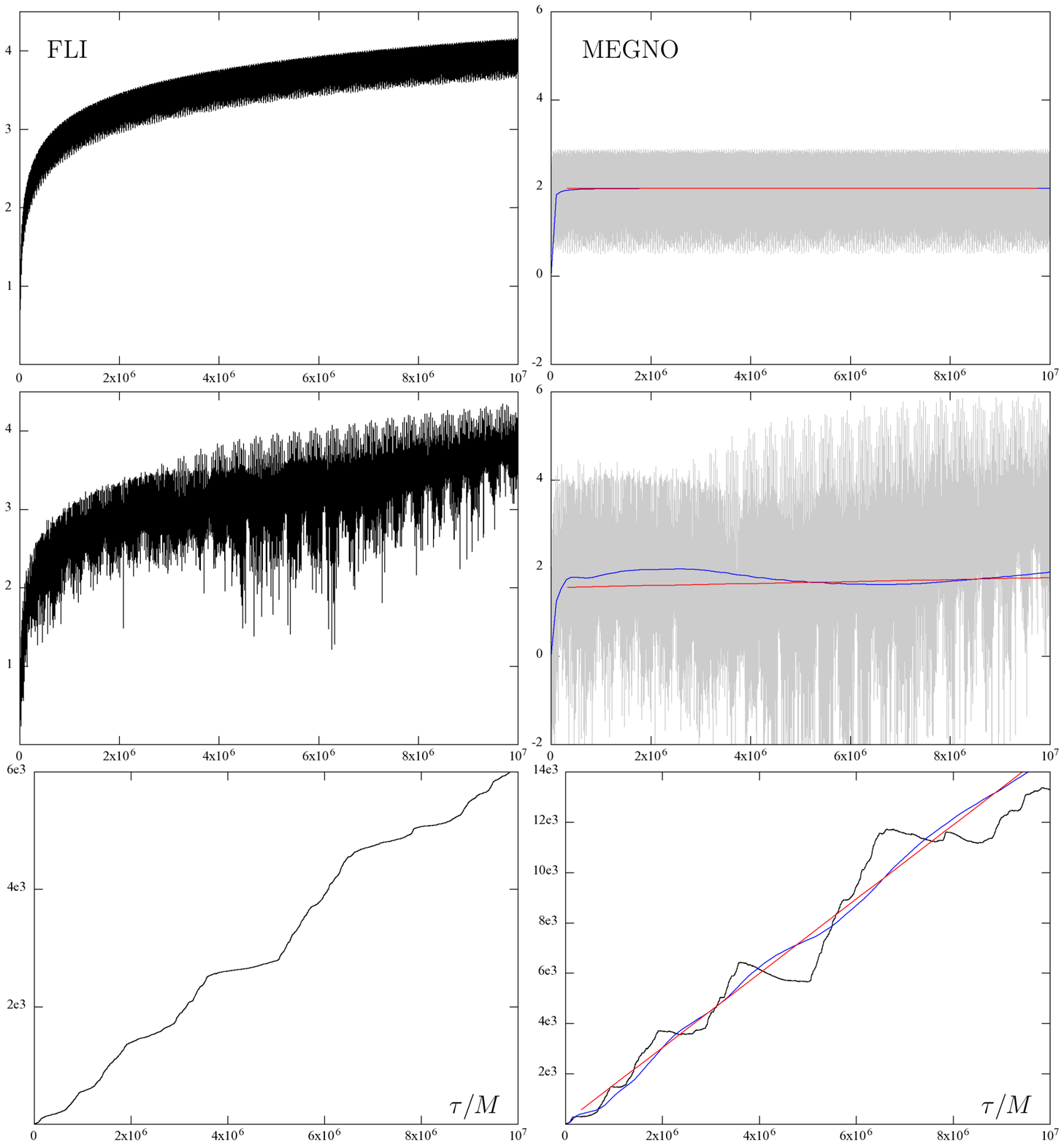}
\caption
{FLI($\tau$) in the left column and $Y(\tau)$ (MEGNO) in the right column for the three orbits (A, B, C) shown in figure \ref{fig-1}. The average MEGNO $\overline{Y}(\tau)$ is drawn in blue and the linear fit of MEGNO is in red. The obtained values of $\lambda_{\rm max}^{\rm MEGNO}$ are $(2.12\pm 0.01)\cdot 10^{-9}M^{-1}$ for the upper row (orbit A), $(4.50\pm 0.08)\cdot 10^{-8}M^{-1}$ for the middle row (B) and $(1.475\pm 0.001)\cdot 10^{-3}M^{-1}$ for the bottom row (C).
\label{fig-4}
}
\end{figure*}

\begin{figure*}
\includegraphics[width=\textwidth]{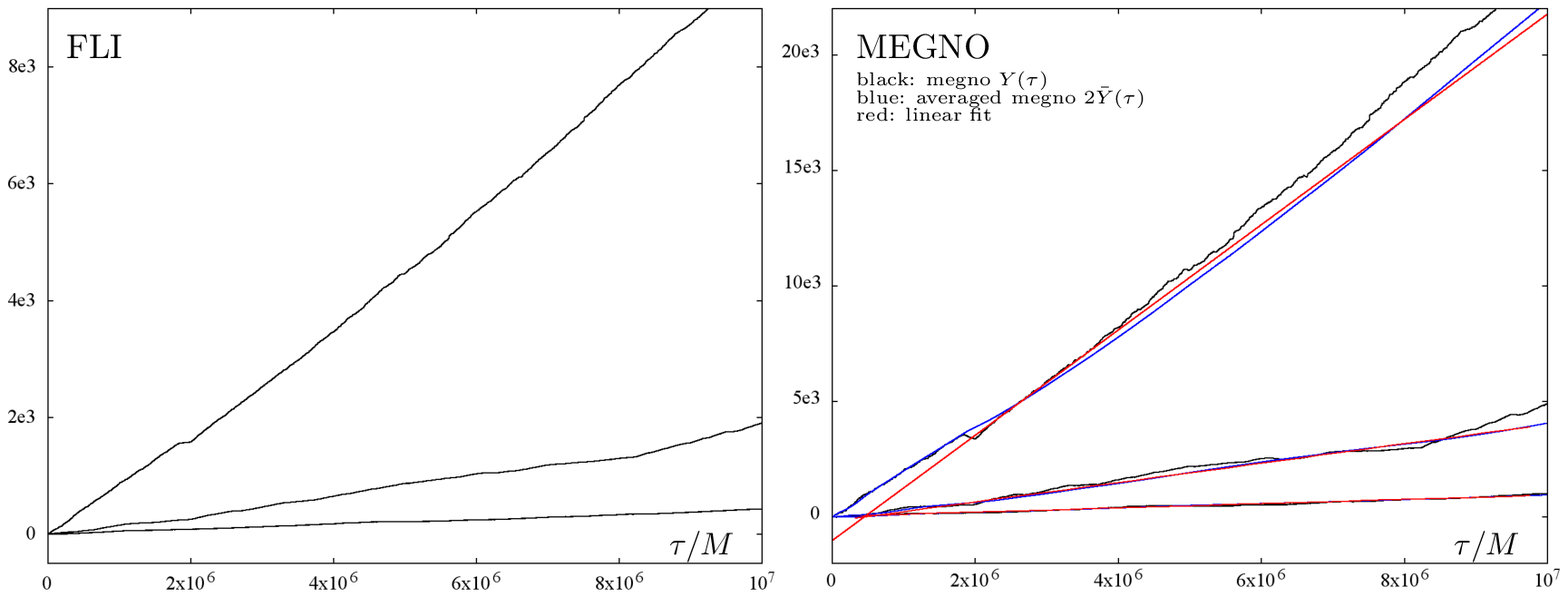}
\caption
{FLI($\tau$) on the left and $Y(\tau)$ (MEGNO) on the right, computed for the three orbits (D, E, F) shown in figures \ref{fig-2} and \ref{fig-3}. The average MEGNO $\overline{Y}(\tau)$ is drawn in blue and the linear fit of MEGNO is in red. The obtained values of $\lambda_{\rm max}^{\rm MEGNO}$ are $(9.172\pm 0.003)\cdot 10^{-5}M^{-1}$ for orbit D (the bottom one), $(4.192\pm 0.001)\cdot 10^{-4} M^{-1}$ for orbit E (the middle one) and $(2.278\pm 0.001)\cdot 10^{-3}M^{-1}$ for orbit F (the top one).
\label{fig-5}
}
\end{figure*}

\subsection{Lyapunov characteristic exponents}

We will compute the mLCE by relation (\ref{mlce}) and adopt the two-particle approach, so we will integrate two geodesics starting from slightly different initial conditions; we set the initial distance between them $|\Delta{\bf x}(0)|$ at $(10^{-7}\div 10^{-9})M$ for motion taking place in the radial range about $10M-100M$, while at $(10^{-4}\div 10^{-2})M$ for motion at much larger radii relevant for our Galactic nucleus (section \ref{galactic-nuclei}). The ratio of the actual to the initial separation is then recorded in some constant step of proper time $\tau$, renormalising the deviation vector to the initial norm $|\Delta{\bf x}(0)|$, together with renormalising the momentum deviation by the same factor, whenever the displacement reaches $10^{-1}M$ (or, in case of larger-scale motions in the Galactic nucleus, we use $10M-100M$ as the threshold).
The integration time has to be sufficient in order to reach a satisfactory convergence of mLCEs. Specifically, we follow the selected orbits for $\tau_{\rm max}=10^7 M$ which corresponds to some $4\cdot 10^4$ cycles about the central black hole. We then plot the behaviour of the so-called time-dependent mLCE
\begin{equation}  \label{mlce_cl_tau}
  \lambda_{\rm max}(\tau)=
  \frac{1}{\tau}\,\ln\frac{|\Delta{\bf x}(\tau)|}{|\Delta{\bf x}(0)|}
\end{equation}
in a log-log scale where it decreases linearly for regular motion while it converges to some non-zero value for chaotic motion. We estimate the final value $\lambda_{\rm max}\equiv\lambda_{\rm max}(\tau\to\infty)$ by an average of the last 5\% of points; for regular orbits this typically represents an upper limit on the actual mLCE value. (We also add a standard deviation of the obtained values, although this represents purely statistical error tied to a given set of points.)

Examples of $\lambda_{\rm max}(\tau)$ together with the corresponding Poincar\'e sections are given in figures \ref{fig-1} and \ref{fig-2}. In figure \ref{fig-1}, three different geodesics (we will call them A, B, C for later reference) with the same global parameters but different initial conditions are shown; the orbit A is regular and belongs to the primary island around the period-one orbit and the orbit B belongs to a regular island with higher period. The behaviour of $\lambda_{\rm max}(\tau)$ is somewhat more noisy for B, but in both cases it converges to zero linearly in the log-log scale, falling down to the value around $1\cdot 10^{-6}M$ after $\tau_{\rm max}=10^7 M$. Regarding that the reciprocal value of mLCE represents a typical time of orbital divergence (the time in which the deviation vector elongates $e$-times), one can just claim, after following the orbits up to $\tau_{\rm max}=10^7 M$, that neither of them is chaotic on the time scale $\approx 10^6M$, which is not very efficient.

The last geodesic (C) in figure \ref{fig-1} fills the whole chaotic layer during the integration time and the final value $\lambda_{\rm max}(\tau_{\rm max})\approx 1.5\cdot 10^{-3}M$ confirms that it is chaotic on time scales of a few orbital periods. Note that the given value of the maximal Lyapunov exponent characterises the whole connected region densely filled with the orbit, but finite pieces of this same trajectory may spend a substantial time in some smaller subsets of phase space, possibly in the vicinity of less unstable periodic orbits or even periodic islands (``sticky" motion), where the character of motion is determined by their stable and unstable manifolds. Symptomatically, after spending some time in a certain region, the orbit may suddenly jump to a different part of the ``chaotic sea" and its local behaviour may change considerably; such a variable behaviour typically produces a ragged profile of $\lambda_{\rm max}(\tau)$, see the bottom row of figure \ref{fig-1}. In order to better visualise this feature, we coloured the points in this chaotic orbit's Poincar\'e section according by their proper time: clearly the colour of chaotic layer is not homogeneous.

Figure \ref{fig-2} presents three orbits (D, E, F) of different levels of chaoticity; orbits E and F have the same global parameters but differ in energy. Orbit D is only slightly chaotic, filling just very thin chaotic layer which originated from a separatrix of the chain of regular islands. As it is well known, such a separatrix, representing a boundary between two different types of motion (e.g. orbits with different periods) and connecting unstable periodic points, is the seed of chaos in the non-linear system.
The thin chaotic layer still contains a chain of smaller periodic islands (see the zoom of the Poincar\'e surface pasted inside the plot), and indeed the orbit yields quite a low value of mLCE, $\lambda_{\rm max}(\tau_{\rm max})=9.88(4)\cdot 10^{-5}M^{-1}$. The second orbit, E, also spends a lot of time in rather narrow layer sticked to periodic islands, but then it sets off into the wide chaotic region around; this behaviour reflects in a somewhat (roughly 4x) higher mLCE value, $\lambda_{\rm max}(\tau_{\rm max})=4.28(6)\cdot 10^{-4}M^{-1}$. The third orbit, F, has higher energy and is strongly chaotic --- it intersects the equatorial plane quite uniformly across the surface of section; correspondingly, the mLCE tends to a much higher value $\lambda_{\rm max}(\tau_{\rm max})=2.250(4)\cdot 10^{-3}M^{-1}$ (note that thanks to the ``clear character" of the orbit, convergence is already reached before $\tau=10^5M$).

\begin{figure*}
\includegraphics[width=\textwidth]{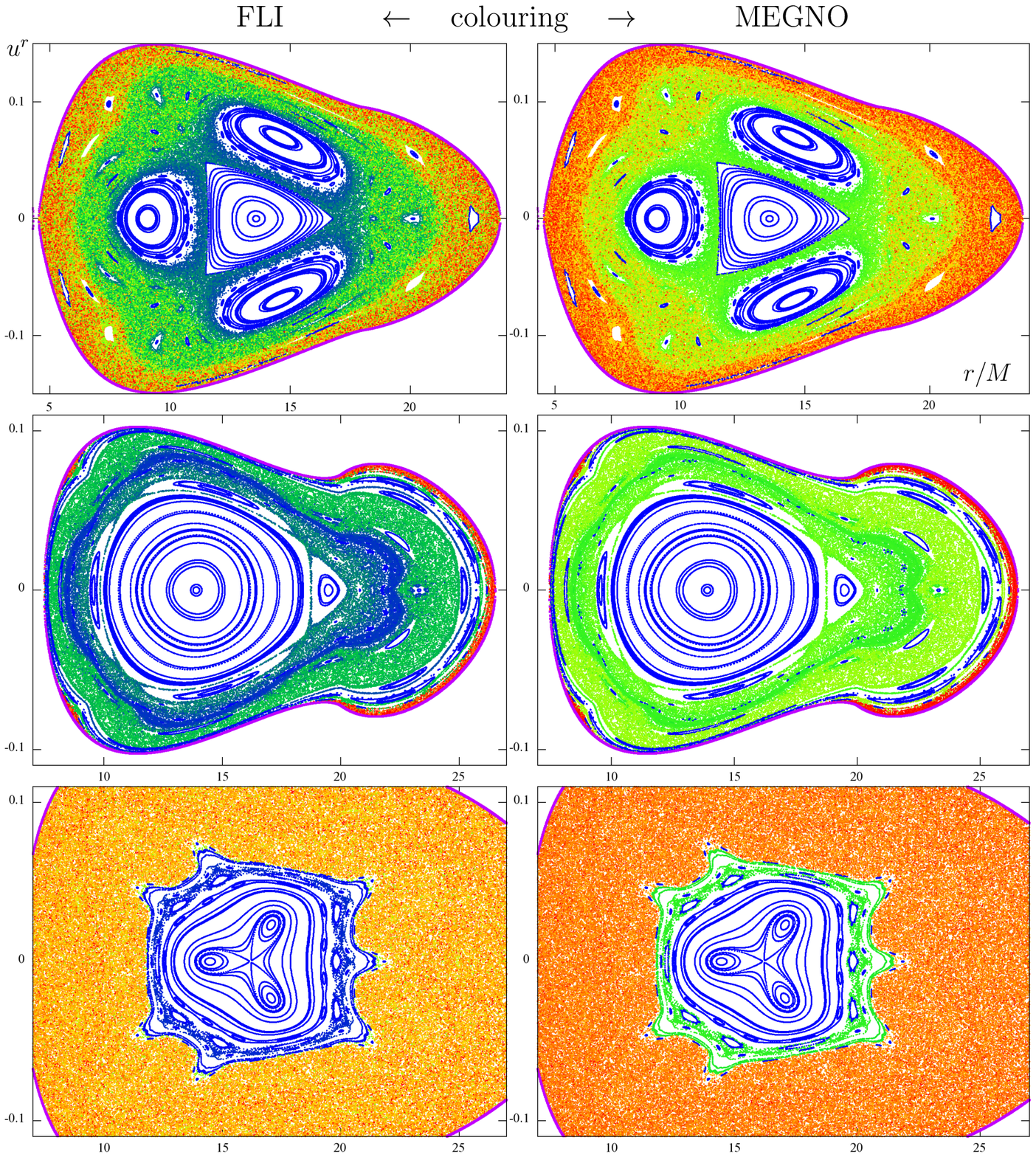}
\caption
{Poincar\'e surfaces of section coloured according to the value of FLI($\tau_{\rm max}$) (left column) and $\bar{Y}(\tau_{\rm max})$ (right column), where $\tau_{\rm max}=250\,000 M$. Choice of parameters: ${\cal M}=M/2$, $r_{\rm disc}=18M$, ${\cal E}=0.9532$, $\ell=3.75M$ in the first row; ${\cal M}=1.3M$, $r_{\rm disc}=20M$, ${\cal E}=0.9365$, $\ell=4M$ in the second row; and ${\cal M}=1.3M$, $r_{\rm disc}=20M$, ${\cal E}=0.941$, $\ell=4M$ in the third row. Whenever $\bar{Y}(\tau_{\rm max})>4$, we add a constant of 200 to it $[\bar{Y}(\tau_{\rm max})\rightarrow \bar{Y}(\tau_{\rm max})+200]$ in order to enhance distinction between regular and weakly chaotic regions. Such an adjusting is possible since the MEGNO indicator approaches a distinct universal value (namely 2) for regular orbits, whereas for chaotic motion it converges towards larger values (typically of the order of hundreds to thousands). This is its main advantage over the FLI, while otherwise both columns are seen to provide the same information. The colours going from blue to red in the visible-spectrum order correspond to FLI increasing in the range 0--350 (left) and to MEGNO increasing in the range 0--500 (after adding 200 to every value above 4, which makes the blue of regular islands more contrasting).
\label{fig-6}
}
\end{figure*}

\begin{figure*}
\includegraphics[width=\textwidth]{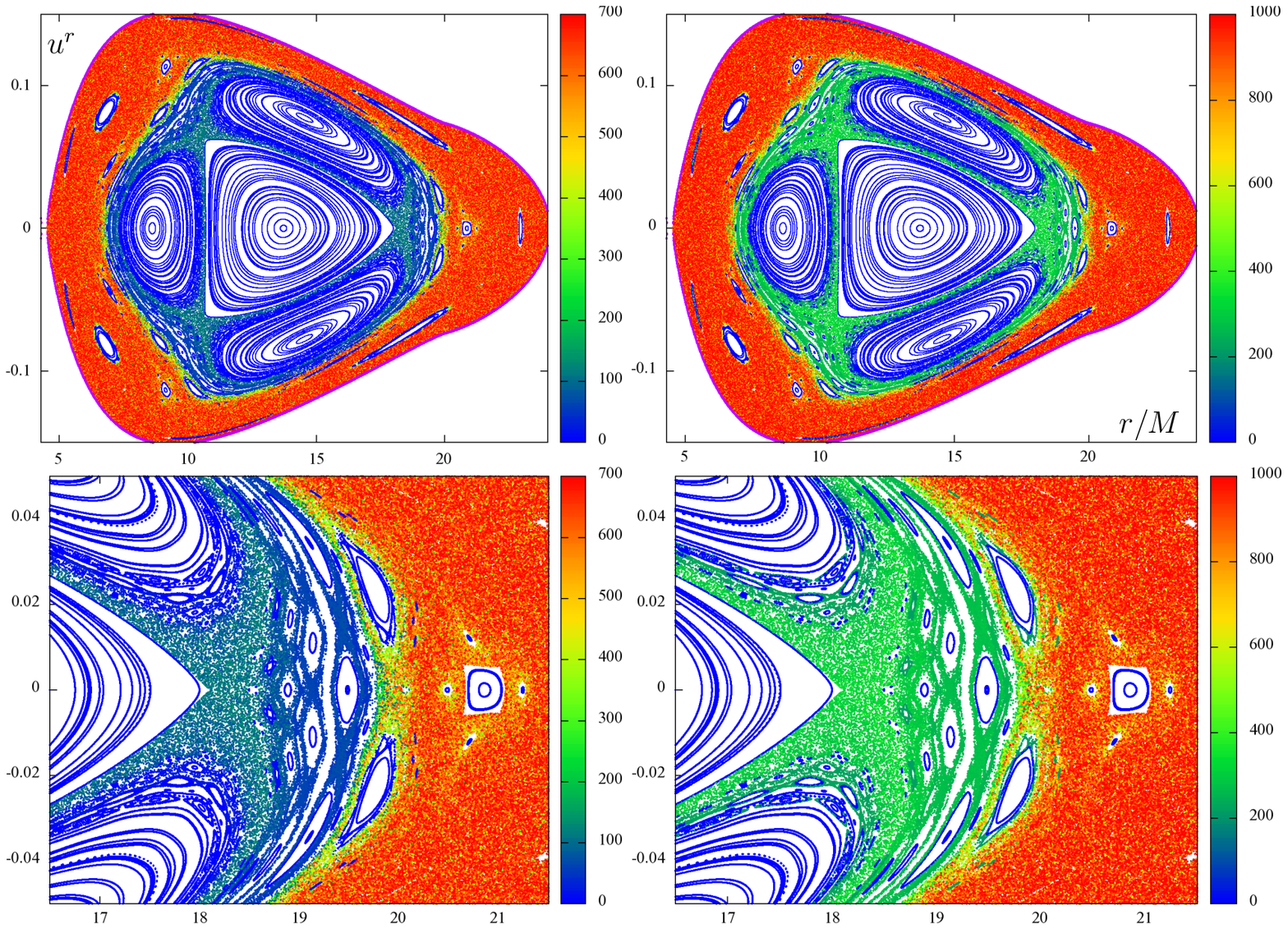}
\caption
{Poincar\'e surfaces of section coloured according to the value of FLI($\tau_{\rm max}$) (left column) and $\bar{Y}(\tau_{\rm max})$ (right column), where $\tau_{\rm max}=10^6 M$. Zoom of the most complex regions is added in the second row. Parameters are chosen as ${\cal M}=M/2$, $r_{\rm disc}=20M$, ${\cal E}=0.955$, $\ell=3.75M$. All values $\bar{Y}(\tau_{\rm max})>4$ are increased by 200 in order to better distinguish between regular and weakly chaotic regions.
\label{fig-7}
}
\end{figure*}

\subsection{Fast Lyapunov indicator}
\label{sec_FLI}

After illustrating that the mLCE behaves in an expected way for our system and, in particular, that it however requires to follow the trajectories for a very long time, let us now focus on the FLI which does not require such a long integration.
First, in figures \ref{fig-4} and \ref{fig-5} we compute the FLI($\tau$) according to relation (\ref{FLI_def}) for the above long trajectories. As expected, FLI remains below 5 for the first two regular orbits, whereas for the remaining chaotic orbits it exceeds several thousands.

We now reduce the integration time to $\tau_{\rm max}=2.5\cdot 10^5 M$, which allows to consider more orbits with different initial conditions and to cover the whole phase space. Three examples of such a map are given in figure \ref{fig-6}, left column. Each of the sections shown there includes a few hundred different trajectories, which are coloured according to the final value FLI($\tau_{\rm max}$). In the first row, the same global parameters are chosen which were used for the orbits A--C in figure \ref{fig-1}. The most chaotic trajectories of this diagram reach some FLI $\approx 300$ and are clearly distinguished from the regular ones which yield FLI $\approx 3$. However, some of the weakly chaotic orbits reach only FLI $\approx 10$, so the distinction is quite uncertain for them. An interesting feature of the first map is the obvious split of the chaotic sea into subregions with different FLI($\tau_{\rm max}$), though we saw in figure \ref{fig-1} (orbit C) that the whole region is a connected chaotic layer. This can be ascribed to the shorter integration time, namely not long enough for some orbits to leave the neighbourhood of periodic islands, and it nicely illustrates the different {\em local} behaviour of particles within the chaotic domain.

The second and third rows in the left column of figure \ref{fig-6} display geodesics having the same parameters as orbits E and F in figure \ref{fig-2}, they only differ in energy. In both cases, the phase-space structure is quite complicated. The second-row map (lower energy) shows the emergence of chaotic layers inside invariant tori, a very thin strongly chaotic region at the periphery of the accessible lobe which, however, almost borders with several higher-period regular islands (this reminds that increasing the ``distance" from the central regular island does not necessarily lead to stronger chaos).
The bottom left map (higher energy) is different: regular motion is concentrated at the centre, surrounded by a large chaotic sea with quite uniformly distributed points and values of FLI($\tau_{\rm max}$). But neither this configuration is ``stable", as inside the primary island there arises a period-three island which for higher energy breaks the surrounding tori.
The last two left diagrams of figure \ref{fig-6} indicate that increasing the particle energy not only enlarges the accessible region but also potentiates the disc's influence on the dynamics. However, as already observed in previous papers, one cannot simply say that higher energy favours chaos; the latter is certainly true for a low end of the energy range, but the energy dependence is rather complicated in general, with abrupt changes around certain values, and rather typically with a regression to regularity for very large values.

\subsection{Mean exponential growth factor of nearby orbits}

In this section we compare the results provided by the FLI and MEGNO coefficients, as obtained from the time behaviour of FLI($\tau$) using the relation (\ref{MEGNO_def}). We also estimate the mLCE from linear regression of $\bar{Y}(\tau)$, on the basis of the relation $Y(\tau)\approx\lambda_{\rm max}(\tau)\,\tau$; let us denote the mLCE determined in such a way $\lambda^{\rm MEGNO}$.
Note that if using the {\em averaged} value of MEGNO $\overline{Y}(\tau)$ for the regression, one has to bear in mind that the averaging decreases the slope of a linear function by a factor of 2. In other words, in order to get a ``smooth" value of MEGNO corresponding to time $\tau$, we have to multiply the value $\overline{Y}(\tau)$ given by relation (\ref{prum_MEGNO}) by 2 before performing the regression.

As already stressed above, the main advantage of MEGNO is a clear distinction between chaotic and regular motion, based on the fact that $\bar{Y}(\tau\to\infty)=2$ for regular orbits regardless of the system details, and also a rather good convergence. When implementing a routine check, one chooses a certain ``order-chaos" threshold $\mu$ which certainly has to be bigger than 2 (in order to include oscillations about the asymptotic value), but typically just a slight shift is sufficient. For a given $\mu$, one can estimate how much chaotic the motion has to be in order to be able to detect its chaoticity within a given time interval. Namely, since the mLCE is deduced from the average slope of MEGNO($\tau$), one finds $\lambda_{\rm max}(\tau_{\rm max})\approx\mu/\tau_{\rm max}$ as a rough bound on how strong chaoticity can be captured. In our computations we set $\mu\equiv 4$ and $10^5\leq\tau_{\rm max}\leq 10^6 M$, hence we are able to ``see" chaos with $\lambda_{\rm max}(\tau_{\rm max})\,{\scriptscriptstyle\gtrsim}\,10^{-5}M^{-1}$.

In figures \ref{fig-4} and \ref{fig-5}, we plot $Y(\tau)$ and $\bar{Y}{(\tau)}$ (or $2\bar{Y}{(\tau)}$) for our selected regular and chaotic trajectories, together with the linear fit which yields the value of mLCE. The estimates of $\lambda^{\rm MEGNO}_{\rm max}$ for the regular orbits are in the range $10^{-8}\div 10^{-9}$, which is about two orders of magnitude below the values of $\lambda_{\rm max}$ obtained from the definition (\ref{mlce}). This indicates better convergence of the present method: the values comparable to those provided by direct computation (with $\tau_{\rm max}=10^7 M$) are reached now at $\tau_{\rm max}\approx 250\,000 M$ already. Note also that the resulting mLCEs of the chaotic trajectories are in a very good agreement for both methods, they differ only by 5\% or less. The results are summarised in table \ref{tab-1}.

\begin{table}
\begin{tabular}{c|c|c|c}
Orbit & $\lambda_{\rm max}(\tau_{\rm max})$   & $\lambda^{\rm MEGNO}$   & $K_2$ \\
A     & $9.35 \cdot 10^{-7}$    & $2.12 \cdot 10^{-9}$    & $1.74\cdot 10^{-7}$ \\
B     & $8.78 \cdot 10^{-7}$    & $4.50 \cdot 10^{-8}$    & $2.7 \cdot 10^{-7}$ \\
C     & $1.405\cdot 10^{-3}$    & $1.475\cdot 10^{-3}$    & $1.23\cdot 10^{-3}$ \\
D     & $9.88 \cdot 10^{-5}$    & $9.172\cdot 10^{-5}$    & $2.97\cdot 10^{-5}$ \\
E     & $4.28 \cdot 10^{-4}$    & $4.192\cdot 10^{-4}$    & $3.12\cdot 10^{-4}$ \\
F     & $2.25 \cdot 10^{-3}$    & $2.278\cdot 10^{-3}$    & $1.69\cdot 10^{-3}$
\end{tabular}
\caption{The values of $\lambda_{\rm max}(\tau_{\rm max})$ and $K_2$ for the orbits from figures \ref{fig-1} and \ref{fig-2}, listed in the same order. Total integration time is $\tau_{\rm max}=10^7 M$. The values were obtained by three different methods, the direct computation from definition, the estimate from linear regression on MEGNO and the estimate from $K_2$ provided by the recurrence analysis.
\label{tab-1}
}
\end{table}

Naturally we compare now the figures where FLI and MEGNO have been used for colouring of the passage points in Poincar\'e diagrams. We proceed similarly as in section \ref{sec_FLI}, recolouring the same diagrams as there according to the achieved MEGNO values $\bar{Y}(\tau_{\rm max})$; the maps obtained are shown in the right column of figure \ref{fig-6}. Taking advantage of the universal small value of MEGNO for regular orbits ($\bar{Y}(\tau\to\infty)=2$), which makes the distinction between regular and chaotic motion more reliable, we set the threshold at $\mu=4$ and whenever $\bar{Y}(\tau_{\rm max})$ is found to be bigger than 4, we add 200 to its value (the added constant is optional, of course). This makes the difference between weakly chaotic and regular orbits more clear visually. (Without this adjustement, the plots would look almost the same as those coloured by FLI, just with a slightly different value range.) Figure \ref{fig-6} demonstrates that both the FLI and MEGNO capture the main features of the dynamics, with fine details better revealed by the latter thanks to having emphasised the order-chaos transition by the additional factor of 200. This is well visible in the second and third rows, where some orbits near the regular islands look rather regular in the left column, while on the right the adjusted MEGNO values reveal their chaotic nature.

In order to further support the above conclusion, we have computed another multi-orbit map spanning a significantly longer time $\tau_{\rm max}=10^6 M$; it is depicted in figure \ref{fig-7} together with a zoom of the most complicated part of the phase space. The section coloured according to the adjusted average MEGNO well reveals fine structures like small periodic islands embedded inside thin chaotic layers, while the colouring by FLI (which does not enable a similar adjustement) does not render such a resolution (and the longer integration time does not improve it much).

\subsection{On visualisation of chaos: animations}

Chaotic dynamics yields lots of nice geometry, but it very difficult to illustrate, within finite space, its most interesting aspect, namely the {\em chaotic} dependence on parameters. Actually, a short series of static plots can indicate an overall tendency, but it is even more remarkable how non-monotonous, irregular these dependences often are, with sudden changes of the phase portrait occurring/disappearing within very narrow parameter ranges. In more complex systems, such features may even remain unnoticed, because numerical scanning of the phase space is performed with a certain finite ``sampling step".

In order to explore more closely the role of our crucial parameters, the relative disc mass and the particle energy, we computed sequences of MEGNO-coloured Poincar\'e maps with gradually increasing relative-mass or energy value while leaving all the other parameters constant, and compiled animations of them. The resulting two videos are available in the online version.
The first animation shows dependence of the phase portrait on particles' energy, for a system with disc mass ${\cal M}=1.3M$ and inner radius $r_{\rm disc}=20M$ and with angular momentum $\ell=4M$. It starts from the value ${\cal E}=0.931$ which is then increased by a step of $0.001$; within the range $0.933\leq {\cal E}\leq 0.939$ the step is only $0.0001$, however, because the phase space changes quite wildly there. The accessible phase lobe emerges near the radius $r=11M$ and contains just completely regular motion at first. Until ${\cal E}=0.9331$ it grows as the particle can move regularly in larger and larger region. At ${\cal E}=0.9332$ another lobe arises at higher radii, namely where the disc's density has its maximum; the motion in this new lobe is also regular. With increased energy the regular islands with higher period arise inside the primary island. First chaos is seen at ${\cal E}=0.9337$, near the periphery of the secondary lobe. This secondary lobe quickly gets strongly chaotic, as indicated by very high values of FLI achieved immediately on the next snapshot (${\cal E}=0.9338$). On the following one (${\cal E}=0.9339$), the two lobes merge and the chaos breaks over the periphery of the primary lobe. The structures from the prior two lobes begin to merge, while the chaos recedes (the orbits' FLI decreases) surprisingly, and new regular tori appear which encircle both the former primary islands and small higher-periodic islands. The strongly chaotic peripheral region remains almost untouched during this process, but when the energy reaches about 0.9362, it gets rather narrow and at ${\cal E}=0.9365$ many regular islands begin to emerge within the primary island. This however brings new resonances which, as the new structures travel towards the edge of the lobe, give rise to a large chaotic sea there. At ${\cal E}=0.958$ a channel in the potential well opens and the particles can fall into the black hole. For very high values of energy, the central island usually restores and spreads, as the dynamics gradually returns to a rather regular behaviour.

The second animation illustrates the dependence on relative mass of the disc with respect to the black-hole mass, ${\cal M}/M$, for $r_{\rm disc}=20M$, ${\cal E}=0.947$ and $\ell=4M$. The overall tendency of evolution of the phase space is quite similar, there also first appears the primary lobe, then the secondary one around the disc's density maximum, the lobes merge and higher-periodic islands accompanied by separatrices breaking then into chaotic layers arise and shift towards the peripheral region in a complicated order. The dynamics is strongly chaotic when ${\cal M}\sim M$, while for very high values of the disc mass it rather returns to a more regular pattern.

\section{Comparison with previous results and other methods}
\label{comparison-with-previous}

It is interesting to compare the mLCE estimate with the value of the second-order R\'enyi's entropy $K_2$ (it reflects a certain type of correlation in the system). The latter is one of the quantifiers of chaos which can be computed from the statistics of orbital recurrences to prescribed domains in the phase/configuration space or in their subspaces; we presented the recurrence analysis of our system in previous paper \citep{SemerakS-12}. The R\'enyi's entropy is given by a slope of the cumulative histogram of diagonal lines of the recurrence plot and represents a lower estimate on the sum of positive Lyapunov exponents. Note that there is an important difference between recurrence quantifiers and the coefficients of orbital divergence discussed here: for obtaining LCEs etc. one has to compute two nearby trajectories (or solve the variational equation along the flow), which requires to know the underlying equations of motion, whereas the correlation entropy $K_2$ and similar quantifiers can be estimated from the recurrence analysis of just partial (e.g. observational) data containing only very restricted information about the system, for example, from a time series of only one dynamical variable (e.g. one coordinate). Although the computation of $K_2$ is quite tricky and requires some experience with recurrence analysis and tuning of its parameters, the values of mLCEs found in the present paper well agree with the estimate obtained from $K_2$ in the previous one. More specifically, the mLCEs computed from the MEGNO slope are slightly larger than $K_2$ for chaotic orbits C--F, while they are smaller than $K_2$ for regular orbits A, B (for which the convergence of MEGNO is better than the direct computation of $\lambda_{\rm max}$). The comparison thus confirms the results presented previously.

A more systematic test of the indicators derived from orbital-deviation evolution was carried out by \cite{Lukes-G-etal-08}; their comparison with spectral methods has recently been performed by \cite{MaffioneDCG-13}.
Here we select figure \ref{fig-2} for such a comparison and add figure \ref{fig-3} where the same three orbits are represented (i) by power spectra of the vertical coordinate $z=r\cos\theta$, denoted $P_\omega(z)$, and (ii) by the averaged ``quadratic deviations from random walk", computed from the added-up directions in which the geodesics recurrently pass through the prescribed phase-space boxes, $\bar\Lambda(\tau)$. This second method was originally designed by \cite{KaplanG-92} to distinguish between deterministic and random evolutions, but we learned in the preceding paper that it can be very efficient ``on the other side of the spectrum", namely in recognising tiny differences between the amount of chaos in {\em very weakly} perturbed systems (or rather between very weakly chaotic orbits or their parts). The rate of recurrent vanishing of a preferred local transit direction is in our case computed in a phase space spanned by a time series of the particle's $z(\tau)$ location and by its two time-delayed copies $z(\tau-\Delta\tau)$ and $z(\tau-2\Delta\tau)$. The result depends on $\Delta\tau$ in a way that is specific for different degrees of chaoticity (or even stochasticity), while the asymptotic {\em value} (approached for $\Delta\tau\rightarrow\infty$) itself brings a clear information: namely, for regular motions $\bar\Lambda$ keeps close to unity, for chaotic motions it decreases to some smaller value and for random noise it falls towards zero.

This is an advantage over power spectra which are often hard to classify (as ``more or less chaotic"), as also illustrated by figure \ref{fig-3}. It well agrees with its counterpart \ref{fig-2} where Poincar\'e diagrams and mLCEs of the same orbits were displayed. Let us point out that $\bar\Lambda(\Delta\tau)$ dependences are plotted only for three selected short intervals of time delay rather than in a full available range as in preceding paper. It is worth to note the difference between the second and the third rows: in figure \ref{fig-2} as well as according to the left column of figure \ref{fig-3} (power spectra), the third-row orbit (F) is more chaotic than the second-row one (E), the only exceptions being the mLCE value around $\tau\sim 2000M$ and the low-frequency part of spectra. Correspondingly, the value of the Kaplan \& Glass quantifier $\bar\Lambda$ (right column of figure \ref{fig-3}) first decreases considerably more slowly for the E orbit, but for large $\Delta\tau$ it even gets {\em below} the value produced by the ``chaotic-sea" F orbit. This confirms that orbit E involves more correlation on short time scales, but on larger scales the correlation washes away; this is consistent with the orbit's time evolution: we checked that in figure \ref{fig-2} it first circumscribes the regular islands (blue), then sticks to them for some time (green) and only then spreads out to a large chaotic region (red). Note that in figure \ref{fig-2} the maximal integration time is $\tau_{\rm max}=10^7 M$, while in figure \ref{fig-3} only the first $10^6 M$ of orbital evolution is included.

It is also interesting to subject to such a comparison the quantifiers provided by the recurrence analysis. Namely, the recurrence properties are somewhat different, ``more basic" in the sense that they can be inferred just from a (discrete) time series, actually even from a series bearing just partial information (like {\em one} coordinate), whereas the study of orbital deviation requires the knowledge of equations of motion (plus the evolution of transversal perturbation). We can, for instance, directly compare top row of figure 12 in the preceding paper \citep{SemerakS-12} with top row of figure \ref{fig-6} in the present one; they both show hundreds of geodesics characterised by energy ${\cal E}=0.9532$ and angular momentum $\ell=3.75M$, scanning the phase space of the field of a black hole ($M$) surrounded by a disc with mass ${\cal M}=M/2$ and inner radius $r_{\rm disc}=18M$. (Note that three particular trajectories occurring in these plots are shown in figure \ref{fig-1}.) Poincar\'e diagrams in figure 12 of the preceding paper are coloured according to the value of $DIV$ (left) and according to the second Renyi's entropy $K_2$ (right), the former being one of the simplest quantifiers of the recurrence analysis (it is the reciprocal value of length of the recurrence-plot longest diagonal), while the latter is a more sophisticated correlation indicator determined by the slope of the diagonal-line histogram. Poincar\'e diagrams in top row of figure \ref{fig-6} of the present paper are coloured according to the value of FLI (left) and MEGNO (right). Abstracting from the somewhat different colour tones chosen, all the four plots apparently carry the same information. One possible message which stems from this comparison is a further confirmation of efficiency of the simplest recurrence quantifier $DIV$.

\section{A specific astrophysical system: Galactic nucleus}
\label{galactic-nuclei}

We have studied the geodesics in the black-hole--disc/ring system purely as a theoretical dynamical system up to now, choosing its parameters without much regarding our original astrophysical motivation. Let us focus now on ``realistic" situations and ask whether pure gravity can incite a noticeable chaos in actual astrophysical systems, though approximated by a highly symmetric configuration. The given arrangement of sources may exist in galactic nuclei, namely those where a supermassive black hole is ``perturbed" by an accretion disc or a ring. The current models of galactic nuclei suggest that there typically occurs an inner hot accretion disc, having radius of light days at most, not very heavy but perhaps reaching down to the vicinity of the black-hole horizon, and a much more massive molecular torus at much larger radius (50 light years, say). If the environment is sufficiently rarefied so that the physical interaction with gas and radiation is negligible, the motion of an individual star can be approximated by a geodesic in the gravitational field dominated by the black hole plus possibly the outer torus and perturbed by the accretion disc. The physical interaction is more probably relevant in nuclei displaying high activity (which is being ascribed to the presence of matter and fields interacting with the black hole), so we should rather refer to nuclei of ``normal" galaxies like Milky Way.

Yet the nucleus of our galaxy is a complex region with a number of components. Considering only those which are the most important gravitationally, observations indicate the presence of a black hole with mass $M\doteq 4.3\cdot 10^6\,M_\odot$ and of {\em two} molecular-gas and dust tori (usually called circumnuclear rings, CNR), one on radii $r_{\rm ring}\sim(1.2\div 3)\,{\rm pc}$ with mass ${\cal M}\sim(10^4\div 10^6)\,M_\odot$ and the other around the radius $r_{\rm ring}\sim(60\div 100)\,{\rm pc}$ with mass ${\cal M}\sim 3\cdot 10^7\,M_\odot$.\footnote
{Whereas the Sgr A$^\ast$ black-hole mass appears to be quite well settled, the parameters of other nuclear constituents are less certain and a considerable spread in the values of their parameters occurs in the literature. For the inner-ring mass, recent estimates range from $1.2\cdot 10^4\,M_\odot$ \citep{Requena-Torres-etal-12} via $(2\div 5)\cdot 10^5\,M_\odot$ \citep{OkaNKT-11} to $1.3\cdot 10^6\,M_\odot$ \citep{Montero-Castano-etal-09}; \cite{HaasSV-11} adhered to the upper estimate ${\cal M}=0.3M$ when approximating the inner circumnuclear toroid by a thin ring. A thorough review of the content of the Galactic innermost region with a list and discussion of observational values has been given by \cite{Ferriere-12}. The parameters of the outer ring have notably been estimated by \cite{Molinari-etal-11}. Both rings roughly follow the galactic plane, but the outer one appears to be twisted (not planar).}
In terms of the black-hole parameters, the inner ring rests around the radius $r_{\rm ring}\sim 10^7\,M$ and has mass ${\cal M}\sim M/20$, while the outer ring is around the radius $r_{\rm ring}\sim 4\cdot 10^8\,M$ and has mass ${\cal M}\sim 10M$. In comparison with the cold tori, the hot inner accretion structure around the black hole is very small (it reaches up to few hundreds or thousands Schwarzschild radii of the central hole) and bears just small fraction of the black-hole mass. Besides that, the Galaxy centre is also inhabited by stars, black holes and gas; in the central parsec, $(10^6\div 10^7)\,M_\odot$ of such mass is supposed to be present (plus the central black hole). In other galaxies these proportions may be considerably different, however; for (extreme) example, the NGC 1277 galaxy appears to boast a black hole with $M\sim 17\cdot 10^9\,M_\odot$ (thus with more than 10\% of the whole-galaxy mass) in the centre.

We will examine which of the above structures could have some effect on particle dynamics. Approximating the centre by a static (originally Schwarzschild) black hole, the tori by the Bach-Weyl ring solution and the inner accretion disc by the inverted first Morgan-Morgan disc solution, we will ask whether some of these sources surrounding the black hole can perturb the motion of an orbiting star (treated as a test particle characterised just by mass) so as to become chaotic. The star is supposed to orbit above the inner accretion zone but below the circumnuclear rings. Note that we do {\em not} take the other stars into account, describing the test-star orbit as a time-like geodesic in the space-time generated by the black hole plus the surrounding ring(s) or disc alone. (Strictly speaking, the evolution of nuclear star cluster should be studied using the relativistic N-particle celestial mechanics, kinetic-theory approach or orbital-perturbation theory, for example, but the effect of ``the other" stars could also be approximated by a spherical or other potential, which might be incorporated in our system quite easily. This is one of our future plans.)

Before turning to numerical results, let us at least briefly refer to the recent high activity focused on the unique black-hole laboratory of our Galactic centre \citep{MorrisMG-12}. Besides the above gaseous and dusty structures, the main attention is devoted to stars, in particular those which orbit very close to the centre \citep{GenzelEG-10}. One reason is of course testing of general relativity and of the precise nature of ``our" supermassive black hole \citep{MerrittAMW-10,SadeghianW-11,AngelilS-11}, while other questions are evolutionary: where and how the observed stars were born and how did they get to their present orbits \citep{MadiganLH-09,FujiiIFM-10,HaasSV-11}. Both tasks require to evaluate properly various possible ``perturbations" acting on the star motion \citep{Iorio-11,Psaltis-12,AntoniniM-13}. This is also our aim here, but rather than perturbations coming from fine details of the very centre (black hole plus the inner cluster) we are interested in the effect of larger circumnuclear structures (cf. \citealt{PeretsHA-07,Chang-09}, for example), namely the two rings of molecular gas and dust.

\subsection{Numerical results}

\begin{figure*}
\includegraphics[width=\textwidth]{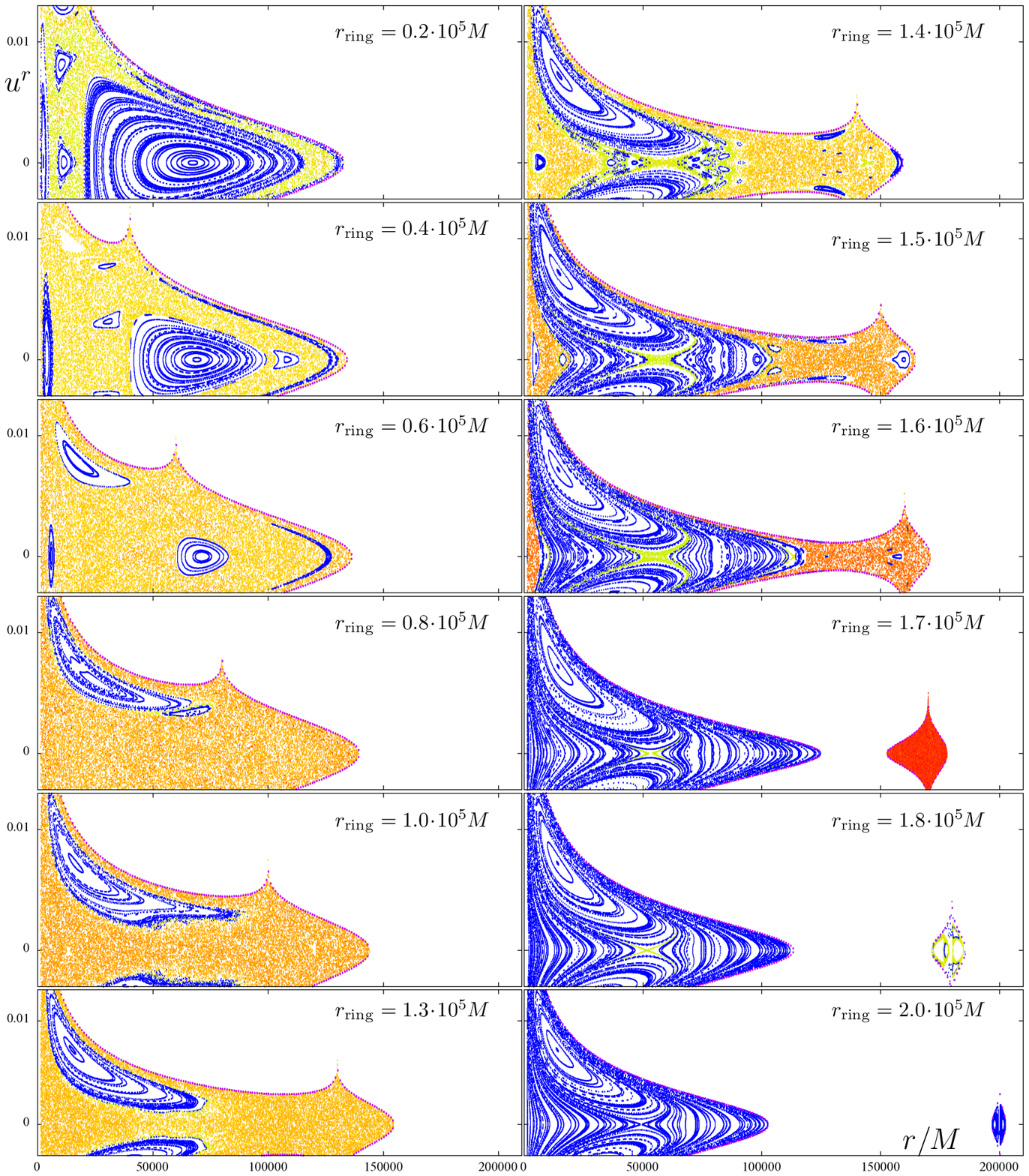}
\caption
{Geodesic dynamics in the field of a Schwarzschild black hole surrounded by a concentric Bach-Weyl thin ring with mass ${\cal M}=M$. Poincar\'e surfaces of section coloured by the value of $\bar{Y}(\tau_{\rm max})$, where $\tau_{\rm max}=15\cdot 10^9 M$, are shown for a sequence of space-times with changing location of the ring: from upper left to lower right (as indicated), the ring's Schwarzschild radius is $r_{\rm ring}=$ 2, 4, 6, 8, 10, 13, 14, 15, 16, 17, 18 and 20 in the units of $10^4 M$. The geodesics have specific energy ${\cal E}=0.999985$ and angular momentum $\ell=50M$. Colours going from blue to red across the visible spectrum correspond to average MEGNO ranging from 0 to 200, but the values above 4 are increased by 200 in order to better distinguish between regular and chaotic regions (thus the range is expanded to 400).
\label{fig-8}
}
\end{figure*}

\begin{figure*}
\includegraphics[width=\textwidth]{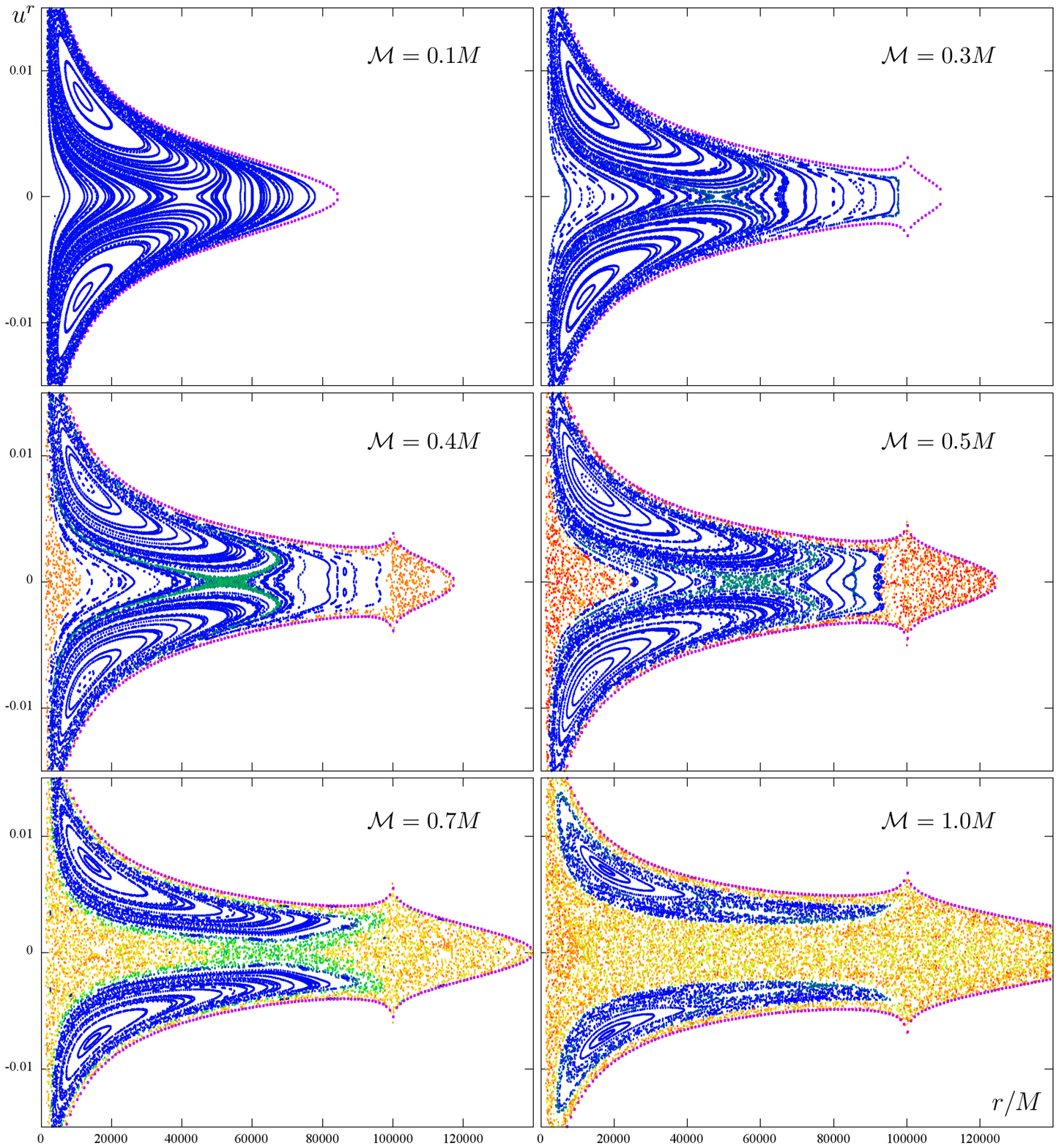}
\caption
{Poincar\'e sections for a sequence of black-hole \& Bach-Weyl-ring space-times with changing mass of the ring: from upper left to lower right, ${\cal M}/M=$ 0.1, 0.3, 0.4, 0.5, 0.7 and 1, while the ring's radius is kept at $r_{\rm ring}=10^5 M$. The geodesics have specific energy ${\cal E}=0.999987$ and angular momentum $\ell=50M$. The sections are coloured by the value of $\bar{Y}(\tau_{\rm max})$, where $\tau_{\rm max}=15\cdot 10^9 M$. Colours going from blue to red across the visible spectrum correspond to average MEGNO ranging from 0 to 150, but the values above 4 are increased by 200 in order to better distinguish between regular and chaotic regions (thus the range is expanded to 350).
\label{fig-9}
}
\end{figure*}

\begin{figure*}
\includegraphics[width=\textwidth]{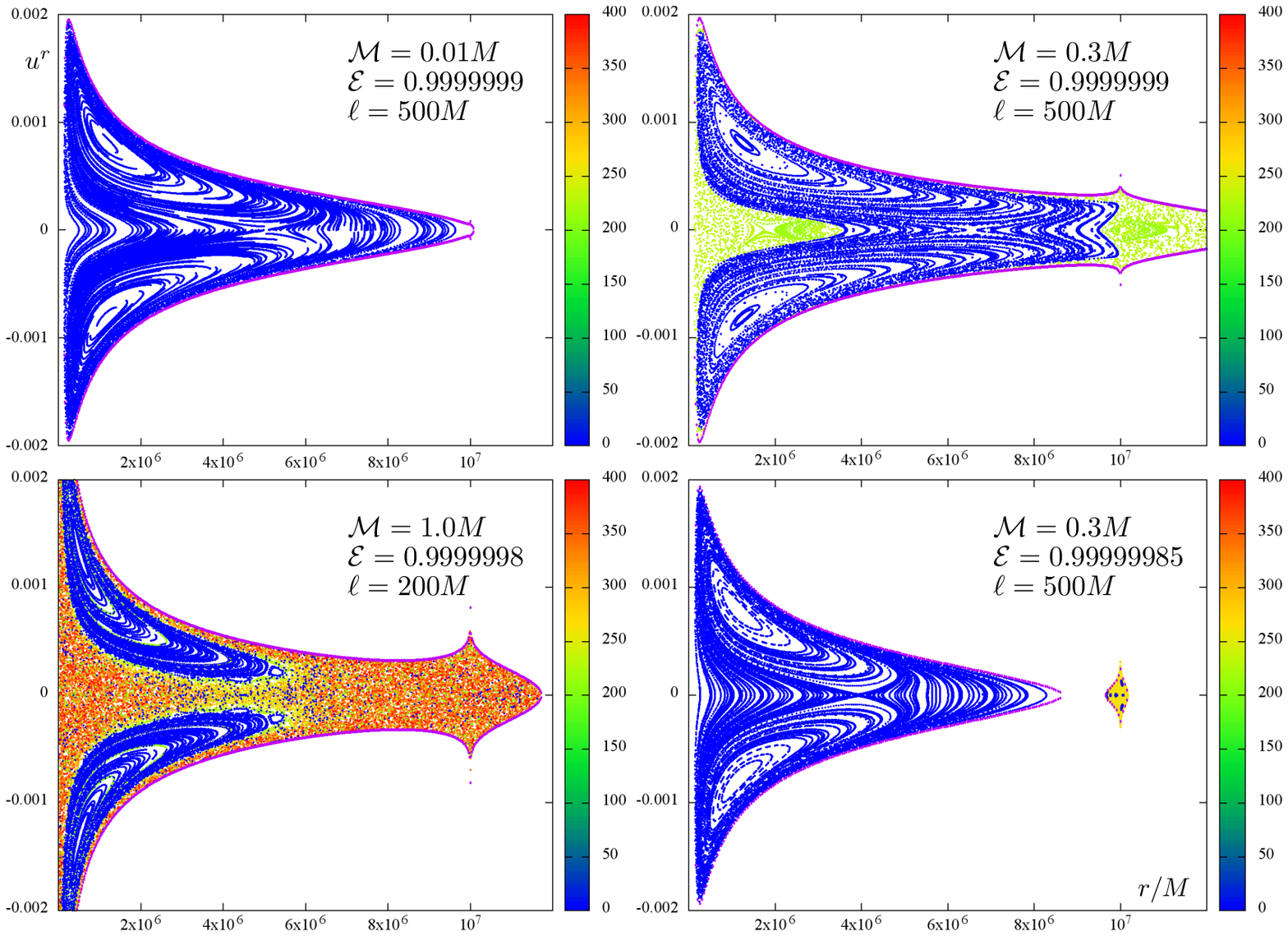}
\caption
{Poincar\'e sections for a sequence of black-hole \& Bach-Weyl-ring space-times with parameters chosen according to the 2-parsec circumnuclear ring in our Galaxy: the ring is on radius $r_{\rm ring}=10^7 M$ and has mass ${\cal M}=0.01M$, $0.3M$ and $1.0M$ as indicated. The first two fields are probed with geodesics having specific energy ${\cal E}=0.9999999$ and angular momentum $\ell=500M$, while for the third (with the least realistic mass ${\cal M}=1.0M$) the geodesic constants are chosen ${\cal E}=0.9999998$, $\ell=200M$. The last plot shows how the motion is considerably more regular when it is confined to the region not reaching up to the ring; the parameters there are ${\cal M}=0.3M$, ${\cal E}=0.99999985$ and $\ell=500M$ in that case (they are the same as in the plot above, just the orbital energy is by $0.00000005$ lower). The points are again coloured by the MEGNO value, with the values above 4 increased by 200.
\label{fig-10}
}
\end{figure*}

\subsubsection{2-parsec ring}

Figures \ref{fig-8} and \ref{fig-9} present a series of equatorial Poincar\'e sections of the geodesic dynamics in the black-hole--ring field; the former illustrates the dependence on ring's location, while the latter illustrates the dependence on its mass. In figure \ref{fig-8}, the mass of the ring is kept at ${\cal M}=M$ and the ring radius is changing in the range $(2\div 20)\cdot 10^4\,M$; therefore, as compared to the parameters of the Milky-Way inner circumnuclear ring, the above ring is about 10 times heavier and lies about 50 times closer to the centre. In other words, the above parameters do {\em not} correspond to actual relations in the Galactic nucleus, but they are also not orders of magnitude different. The total integration time is chosen $\tau_{\rm max}=15\cdot 10^9 M$ which in this case represents some 200 orbital periods. The mass of the central black hole is set at $M=4.3\cdot 10^6 M_\odot$, which fixes our length and time scale according to $1M\doteq 6.6\cdot 10^6{\rm km}\doteq 22{\rm s}$. One orbital period is thus about 50 terrestrial years and the particle/star motion is followed for about $10\,000$ years. The linear orbital velocity ranges from about a hundred to a few thousand km/s.

The figures show that the effect of the ring on the motion of particles is quite pronounced and strongly depending on the ring location with respect to the accessible region. Generically, the ring affects mainly those orbits which pass in its vicinity. For given fixed constants of geodesic motion (energy and angular momentum at infinity), the system thus appears the most chaotic when the ring passes just through the centre of the accessible region. Then the primary island with period 1 shrinks and then disappears completely. For $r_{\rm ring}=80\,000 M$, for example, $\lambda_{\rm max}^{\rm MEGNO}\approx (5\cdot 10^{-8}\div 5\cdot 10^{-9})\,M^{-1}$ which confirms quite strong chaos on the time scale of only a few periods. When the ring position approaches the outer boundary of the lobe, a regular region inside grows in the form of period-two islands. Finally a secondary lobe enclosing just the ring detaches, leaving the primary lobe almost regular (except for a small territory near the separatrix between the two regions); the secondary lobe is heavily chaotic. Shifting the ring still farther, the primary lobe stabilises further, while the chaotic lobe tied to the ring gradually shrinks and becomes more regular as well.

Figure \ref{fig-9} brings a similar sequence which demonstrates a response of the dynamics on the disc mass: the ring is kept on $r_{\rm ring}=10^5 M$ and its mass ${\cal M}$ changes from $M/10$ to $M$. Again, if keeping the geodesics' global parameters (energy and angular momentum at infinity) constant, the dynamics is quite sensitive to the mass parameter. Namely, while varying the mass by a factor of 10, one proceeds from a completely regular to a rather chaotic system. The phase-space structures evolve in a similar way as in the preceding case.

In order to follow the actual Galactic parameters still more closely, we also analysed the case when the ring is placed at $r_{\rm ring}=10^7 M$. Figure \ref{fig-10} brings three plots which mainly differ in the ring mass, ${\cal M}=0.01M$, $0.3M$ and $1.0M$, respectively, and one plot where the particles/stars do not have enough energy to approach the ring. The phase spaces of the $0.3M$ and $1.0M$ cases do contain chaotic regions, but a closer look (plotting of individual trajectories) shows that only those orbits are irregular which get close to the ring. Actually, the chaotic region around the origin is connected, by these orbits and via the narrow peripheral layer, with the chaotic vicinity of the ring. This is why we also added a plot (the last one) where the motion is confined to a smaller region not reaching to the very ring. For such a motion everything is regular. More precisely, there do exist a chaotic region tied to the ring, but this is not connected with the main accessible region (and hence the latter remains regular).

\subsubsection{Accretion disc}

The hot accretion disc in the black-hole vicinity does not appear to have a significant gravitational effect on the dynamics of stars orbiting the centre in the distance range of the well known S-stars or somewhat farther. More specifically, we modeled the accretion structure by the inverted 1st Morgan-Morgan disc with the inner edge on $r_{\rm disc}=100M$ (most of its mass is concentrated close to the inner radius, say between $100M$ and $500M$ --- see e.g. \cite{Semerak-03}) and having mass ${\cal M}=0.01M$. Geodesics with specific energy ${\cal E}=0.99999$ and angular momentum $\ell=50M$, launched in all possible directions and scanning the corresponding accessible phase-space region (extending roughly from $1270M$ to $10^5 M$), proved to be completely regular, so we do not even show the resulting Poincar\'e section nor the respective values of mLCE, FLI and MEGNO quantifiers. As the mass $0.01M$ is highly overshot for the actual inner accretion structure supposed to exist in the Galaxy nucleus, we can conclude that the latter does not destabilise the motion of S-type stars.

\subsubsection{80-parsec ring}

A similar conclusion also applies to the effect of the larger ring resting/orbiting between 60 and 100 parsecs on stars orbiting on smaller radii. We approximated the ring by the circular Bach-Weyl ring with radius $r_{\rm ring}=4\cdot 10^8 M$ and with mass ${\cal M}=10M$, and scanned the phase space with geodesics having energy ${\cal E}=0.99999993$ and angular momentum $\ell=1000M$. All their accessible region, extending from $6\cdot 10^5 M$ to some $2.1\cdot 10^7 M$, proved to be regular. (We again do not attach the resulting figures and indicator values.)

\section{Concluding remarks}

It is interesting and desirable to study chaotic dynamics by various methods, because different of them are sensitive to different features and they also differ in coordinate-dependence \citep{GelfertM-10}. We have subjected to such a study the time-like geodesic dynamics in the static and axisymmetric field of a Schwarzschild black hole perturbed by the presence of a concentric thin disc or ring. After revealing basic tendencies of this system on Poincar\'e surfaces of section and on time-series of one of the phase-space variables and its power spectra (paper I), and after comparing the results with those provided by two effective recurrence methods while focusing on selected ``sticky" orbits in more detail (paper II), we have checked in the present paper whether similar conclusions also follow by computing the basic coefficients describing the rate of orbital divergence, namely the maximal Lyapunov characteristic exponent, the fast Lyapunov indicator and the MEGNO indicator. Regarding our motivation, stemming from astrophysical systems dominated by black holes (in particular galactic nuclei with lower density of interstellar matter), we then set the system parameters at values corresponding to the structures observed/supposed in our Galactic nucleus --- the central supermassive black hole, the inner hot accretion disc around it, and two circumnuclear molecular and dust rings on radii around 2pc and 80pc. Of these, only the smaller circumnuclear ring have been found to be able to partially destabilise the motion of stars (treated as test particles), but only if these can approach it closely.

We have naturally chosen our-Galaxy example for the above order/chaos test, but similar structures (discs, rings, tori) are also being observed in other galaxies, both quiet and active. Since the parameters corresponding to our Galactic nucleus are not {\em that} far from the range where chaos begins to occur, at least when speaking about the smaller circumnuclear ring, it is likely that in some galaxies a similar analysis of orbital dynamics would yield interesting results.

The last remark concerns the validity of exact models (exact solutions of Einstein's equations) we are using to describe the gravitational field. First, the influence of the central star cluster (of ``other stars") should be incorporated. This would be rather easy if the cluster were only approximated by a kind of spheroidal potential. Second, it would certainly be more realistic to consider thick toroids instead of infinitely thin rings to model the circumnuclear galactic structures, at least if stars may get to their close vicinity. This is also a feasible task, though the metric is then much more demanding for practical computations (see e.g. \citealt{SachaS-05} where we presented an example of a static black hole surrounded by a thick toroid). Third, a {\em physical} interaction of stars with circumnuclear environment should also be accounted for somehow, but it is clearly beyond our present scope to discuss it here. Finally, as relativists we would very much care for incorporating {\em rotation} in our black-hole--something system, both for theoretical reasons (non-linear interplay of rotational draggings from black hole and from the disc/ring) and certainly because rotation is ubiquitous in astrophysics. This turned out to be extremely difficult, unfortunately, at least within exact relativistic description; various procedures using ``solution-generating techniques" have failed due to unphysical features present in the results (like in our own attempt described in \citealt{ZellerinS-00,Semerak-02}), while the outcomes of more sophisticated approaches are too complicated for inclusion as a part of some computational scheme, and even for a thorough study of their physical properties (cf. e.g. \citealt{KleinR-05}).

\subsection{Recent results in the field}

Let us also mention several related results which have appeared in the literature recently.
\cite{WangW-11,WuZ-11}
used the FLI to study the dynamics of spinning compact binaries, analysing the effect of terms of different pN orders.
\cite{StachowiakS-11}
proposed a new algorithm for computing the Lyapunov exponents, not based on usual repetitive evolution and rescaling of the connecting vector.
\cite{SeyrichLG-12}
developed a new integration scheme for non-integrable Hamiltonians, by combination the symmetric symplectic integrator with a new step-size controller.
\cite{Lukes-G-12}
demonstrated that the geodesic motion in the Zipoy-Voorhees space-time is non-integrable in general, while
\cite{KovarKKK-13}
came to the same conclusion for the motion of neutral as well as charged particles around the Bonnor's massive magnetic dipole.
\cite{ContopoulosLGA-11} and \cite{BambiLG-13}
analysed simlilarly the geodesic dynamics in the Manko-Novikov space-time and discussed differences versus the Kerr case.
\cite{AlZahraniFS-13}
investigated (generically chaotic) charged-particle motion around a weakly magnetised static non-rotating black hole, being mainly interested in the escape velocity of particles perturbed off the circular orbits.
Very recently \cite{BrinkGH-13}
analysed, in the context of identification of the Sgr A$\ast$ black hole, orbital resonances in the Kerr field in order to estimate where geodesic chaos would occur in case of departure from the Kerr space-time.
Finally, in connection with our plan to include rotation into the black-hole--disc/ring gravitational system and examine how dragging effects change the geodesic dynamics, we should mention that \cite{WangW-12} considered a Kerr black hole mimicked by a pseudo-Newtonian potential and superposed it with a quadrupole halo, asking how the geodesic dynamics responds on spin of the centre and on quadrupole perturbation. They found, among others, that the black-hole rotation rather attenuates the instability, consistently with previous experience acquired in this respect.

\section*{Acknowledgements}

We used the code for geodesic motion in Weyl space-times originally written by M. \v{Z}\'a\v{c}ek
and have been running it on the {\sc Sn\v{e}hurka} cluster of our faculty.
The plots were produced with the help of the {\sc Gnuplot} utility and D. Krause's {\sc bmeps} program.
We thank D. Heyrovsk\'y for interest and comments.
Our work has been supported from the projects GACR-202/09/0772 and MSM0021620860 (O.S.); SVV-267301 and GAUK-428011 (P.S.).

\end{document}